\documentclass[sigconf]{acmart} %
\usepackage{tikz}
\usepackage{amsmath}
\usepackage{tabularx}
\usepackage{booktabs,tabularx}
\usepackage{makecell,pifont}
\usepackage{graphicx} %
\usepackage{capt-of} %
\usepackage{hyperref}
\usepackage{enumitem}
\usepackage{comment}
\usepackage{siunitx}
\usepackage{array}
\usepackage{subfigure}
\usepackage{caption}
\usepackage{subcaption}   
\usepackage{mwe} 
\usepackage{cleveref}
\usepackage{booktabs}
\usepackage[table,xcdraw]{xcolor}
\usepackage[T1]{fontenc}
\usepackage[utf8]{inputenc}
\usepackage{fontawesome5}
\usepackage{mdframed}
\usepackage{stfloats}
\usepackage{balance}

\usepackage{url}                %
\usepackage{xcolor}             %
\usepackage[]{hyperref}  
\usepackage{breakurl}           %

\newcolumntype{C}{>{\centering\arraybackslash}X}           %
\newcolumntype{s}{S[table-format=4,table-number-alignment=center]} %

\usepackage{xurl}

\usepackage{ragged2e}
\newcolumntype{P}[1]{>{\RaggedRight\arraybackslash}p{#1}}

\newcommand{\fakeparagraph}[1]{{\vskip 2pt \noindent\bfseries #1.}}
\usepackage{xcolor}

\newcommand{\res}[2][1]{#2}

\mdfdefinestyle{infoboxstyle}{
  backgroundcolor = gray!12,       %
  linecolor       = gray!50,       %
  linewidth       = 0.6pt,
  roundcorner     = 2pt,
  leftmargin      = 0pt,
  rightmargin     = 0pt,
  innerleftmargin = 8pt,
  innerrightmargin= 8pt,
  innertopmargin  = 6pt,
  innerbottommargin=6pt,
  skipabove       = 8pt,
  skipbelow       = 8pt,
}

\definecolor{ultralightgray}{gray}{0.9}

\newcounter{anec}

\mdfdefinestyle{quoteleft}{
  backgroundcolor=ultralightgray,
  linecolor=red,
  leftline=true, rightline=false, topline=false, bottomline=false,
  linewidth=2pt,
  innerleftmargin=6pt, innerrightmargin=6pt, innertopmargin=2pt, innerbottommargin=2pt,
  skipabove=5pt, skipbelow=0pt
}

\newenvironment{anecquote}{%
  \refstepcounter{anec}%
  \begin{mdframed}[style=quoteleft]\itshape
  \textbf{Anecdote~\theanec. }\ignorespaces
}{%
  \end{mdframed}
}

\AtBeginDocument{%
  }

\copyrightyear{2026}
\acmYear{2026}
\setcopyright{cc}
\setcctype{by}
\acmConference[CCS '26]{Proceedings of the 2026 ACM SIGSAC Conference on Computer and Communications Security}{November 15--19, 2026}{The Hague, Netherlands}
\acmBooktitle{Proceedings of the 2026 ACM SIGSAC Conference on Computer and Communications Security (CCS '26), November 15--19, 2026, The Hague, Netherlands}
\acmDOI{10.1145/3830454.3846651}
\acmISBN{979-8-4007-2871-6/2026/11}
\begin{document}

\title{Keys on Doormats: Exposed API Credentials on the Web} %

\author{Nurullah Demir}
\email{nurullah@cs.stanford.edu}
\orcid{0000-0001-8721-4412}
\affiliation{%
  \institution{Stanford University}
  \city{}
  \state{}
  \country{}
}

\author{Yash Vekaria}
\email{yvekaria@ucdavis.edu}
\orcid{0000-0002-1891-7568}
\affiliation{%
  \institution{University of California, Davis}
  \city{}
  \state{}
  \country{}
}

\author{Georgios Smaragdakis}
\email{g.smaragdakis@tudelft.nl}
\orcid{0000-0002-4127-3617}
\affiliation{%
  \institution{TU Delft}
  \city{}
  \country{}
}
\affiliation{%
  \institution{Stanford University}
  \city{}
  \state{}
  \country{}
}

\author{Zakir Durumeric}
\email{zakir@cs.stanford.edu}
\orcid{0000-0002-9647-4192}
\affiliation{%
  \institution{Stanford University}
  \city{}
  \state{}
  \country{}
}

\begin{abstract} %
API (Application Programming Interface) keys allow applications to authenticate themselves to third-party services. Inadvertent public exposure of these credentials can pose significant consequences, as adversaries can use them to gain privileged access to other services. In this paper, we measure API credential exposure on the web by analyzing 10M~rendered websites. Our findings reveal that API credential exposure on the web is widespread, affecting organizations such as global banks and core infrastructure providers.
We identify 1,748~credentials for accessing 14~providers (e.g., cloud and payment services).
Crucially, we demonstrate that these exposures are largely missed by static analysis. By characterizing web-specific exposure vectors and root causes, we find that 62\% of JavaScript-based exposures manifest exclusively within compiled deployment bundles, while 16\% propagate dynamically through third-party resource inclusions.
Moreover, our longitudinal analysis shows these credentials often persist for months to years. 
We conclude by discussing our responsible disclosure efforts and outlining mitigations to secure web deployment pipelines in the future.

\end{abstract}



\maketitle

\section{Introduction}

Developers use Application Programming Interface (API) keys to authenticate applications to third-party services (e.g., cloud platforms, payment providers)~\cite{sun2022research}. Despite security guidelines advising against hard-coding credentials~\cite{OWASPSMCS, GitHubDocs_NoHardcode, MicrosoftSecretsGuidance2025}, it is common for developers to embed keys in source code or configuration files for convenience~\cite{sinha2015secret}. Unfortunately, if that code is leaked or publicly exposed, secrets can be easily extracted and misused.

Several high-profile incidents highlight the risks associated with exposed credentials and how a single leaked key can lead to data theft, privacy compromise, and reputational damage. For example, Uber’s 2016 data breach was traced to attackers finding an Amazon Web Services (AWS) access key embedded in public code on GitHub~\cite{Uber2017Incident, FTC2018UberRevisedComplaint}. Attackers used the key to access Uber’s cloud infrastructure, exposing private data from 57M~users. Similar exposures have led to data breaches at Slack and The New York Times~\cite{Slack2022SecUpdate, Abrams2024NYTLeak}. %

Typically, organizations attempt to protect themselves by conducting codebase audits using automated tools like {TruffleHog}~\cite{TruffleHog2025} and {Gitleaks}~\cite{Gitleaks}. %
Developer platforms such as GitHub~\cite{GitHubDocs_SecretScanning}, GitLab~\cite{GitLabDocs_SecretDetection}, and Bitbucket~\cite{BitbucketDocs_SecretScanning} have integrated automated scanning, while cloud providers like Google Cloud~\cite{GoogleCloud_DisableLeakedKeys} and Azure~\cite{MicrosoftAzureSecretsNotify} respond to public exposures by automatically disabling leaked keys and/or notifying owners.
Overall, credential exposure remains a systemic challenge. %
Yet the exposure of sensitive service credentials on the web has not been systematically examined.

In this paper, we address this gap by conducting a large-scale {dynamic} analysis of 10M~popular websites, identifying widespread exposed credentials, impacting public-safety systems, critical infrastructure, global banks and healthcare providers. We examine credential persistence and remediation through responsible disclosures, and we analyze the technical and organizational factors that lead to exposure.
Our contributions can be summarized as:

\begin{enumerate}[topsep=0pt]
\item \textbf{Large-Scale Analysis of Credential Exposure.} We conduct the first large-scale dynamic analysis of API credential exposure on the web, identifying 1,748~valid keys to 14~popular services. We uncover credentials belonging to national infrastructure providers and major banks, including a ``Global Systemically Important Bank'' (G-SIB)~\cite{BIS_GSIB_2025} and public-safety systems handling sensitive personal data.

\item \textbf{Root Cause Analysis and Recommendations.} We characterize web-specific exposure vectors, finding that 84\% of exposures manifest within JavaScript resources, including bundled files (62\%) and third-party inclusions (16\%). We show that dynamic analysis is vital for finding exposure; while prior work heavily focuses on analyzing code repositories, the majority of web-based exposures occur only in production environments. Based on empirical evidence and disclosure feedback, we identify the underlying technical and organizational causes, such as insecure build pipelines and environment variable misconfigurations, and provide actionable recommendations for stakeholders. 

\item \textbf{Persistence and Public Disclosure.}  
We show that exposed credentials often persist for months to years (average 12~months), highlighting long-term security risks. However, our large-scale responsible disclosure effort led to the removal of about half of the exposed credentials across analyzed services within two weeks.

\end{enumerate}

\vspace{2pt}
\noindent Our results highlight the continued risk of exposed credentials and the shortcomings of current approaches. 

\section{Background}
\label{sec:background}
We provide the terminology and threat model relevant to our study.
 
\vspace{2pt}
\fakeparagraph{Terminology} \quad
Here, we define the terminology used throughout this paper. A \textit{hostname} refers to a fully qualified host identifier, including subdomains (e.g., {foo.bar.com}). A \textit{domain} refers to the eTLD+1: the registrable portion of a hostname (e.g., {bar.com}). We use the term \textit{website} to denote all hostnames that share the same eTLD+1 domain (e.g., both {bar.com} and {foo.bar.com} are considered part of the same website). A \textit{webpage} denotes a specific URL path or resource (e.g., {https://www.bar.com/test.html}). 

\vspace{2pt}
\fakeparagraph{Threat Model} \quad
\label{sec:threat-model}
We model an adversary that has no privileged access but can collect and analyze public web content. Such an adversary can index, store, and inspect any client-visible resource, including JavaScript files, API responses, or embedded configuration data. The attacker cannot compromise servers, intercept encrypted traffic, or obtain secrets through insider or malware access. 
Our study captures a worst-case exposure scenario, quantifying the prevalence and origin of client-side credential exposure.

\section{Related Work}

Our paper builds on two bodies of prior work:

\vspace{2pt}
\fakeparagraph{Credential Exposure} \quad
Prior research on credential exposure has primarily looked at non-web ecosystems such as GitHub~\cite{farinella2021git,meli2019bad,wen2022secrethunter,feng2022automated,niu2023codexleaks}, cloud~\cite{cable2021stratosphere, springall2016ftp, continella2018there, el2025file},
mobile apps~\cite{zuo2019does,alecci2025evaluating}, VS Code extensions~\cite{liu2025protect}, InterPlanetary File System (IPFS)~\cite{wu2024secrets}, Docker images~\cite{dahlmanns2023secrets}, browser extensions~\cite{chen2026keychaser}, and mini-programs~\cite{zhang2023don}.
Other works have adopted qualitative methods to understand real-world developer practices, such as socio-technical causes of accidental credential exposures~\cite{krause2023pushed} and password-choice habits~\cite{lykousas2023tales}.
Collectively, these studies reveal that credential exposure occurs across diverse ecosystems.  
Most closely related, a recent report by Truffle Security explored credential exposure using static Common Crawl archives~\cite{truffle_blog}. However, by relying only on static WARC files, the approach misses the vast majority of dynamically introduced exposures that do not occur in the initial HTTP response~\cite{commoncrawl_warc}.
Furthermore, it does not distinguish between intentionally public, safe-to-expose keys and genuinely sensitive private credentials, which likely leads to overestimating vulnerability.

\vspace{2pt}
\fakeparagraph{Vulnerability Disclosure}  \quad
Researchers have studied responsible disclosures to address security vulnerabilities such as unintended abuse of server infrastructure~\cite{cetin2016understanding, vasek2012malware}, HTTPS misconfigurations~\cite{zeng2019fixing}, and many others~\cite{li2016you, stover2023website, soussi2020feasibility}.
Over the years, the literature has systematically studied the effectiveness of different notification strategies, including communication channels~\cite{li2016you,stock2016hey,maass2021snail}, message content~\cite{maass2021effective,zeng2019fixing,maass2021best}, sender identity~\cite{cetin2016understanding,maass2021effective}, recipient characteristics~\cite{li2016remedying,ccetin2019tell}, and interdependencies among these factors~\cite{cetin2017make,vekaria2024turning}.
Our research is {not} a vulnerability notification study to understand the effectiveness of different notification controls. 
It is, rather, the first large-scale measurement study of credential exposure on the web.
To this end, we take inspiration from the literature on vulnerability disclosures to devise our responsible disclosure notifications by adopting best practices recommended in these works.

\vspace{2pt}
\noindent \textbf{Results Over Prior Work.}\quad We advance prior work by providing new dimensions to the study of credential exposure:  (i) rather than analyzing static artifacts (e.g., source code repositories) that miss production-only exposures, we conduct a large-scale dynamic analysis of the rendered web, demonstrating that the vast majority of credential exposures manifest exclusively in live production environments---introduced dynamically during build and deployment workflows;   (ii) beyond detection, we quantify how long credentials remain exposed and assess real-world remediation through our responsible disclosure efforts; and (iii) grounded in empirical evidence and disclosure feedback, we identify the technical and organizational root causes of these web-native exposures and highlight their broader implications, including real-world supply-chain risks affecting critical infrastructure.

\section{Methodology}
\label{sec:methodology}

This section describes our methodology for detecting exposed credentials (\Cref{sec:identifying-exposed-credentials}), analyzing persistence (\Cref{sec:methodology-longitudinal}), and conducting responsible disclosure (\Cref{sec:methodology-responsible-disclosures}).

\begin{figure}
    \centering
    \includegraphics[width=\linewidth, height=\textheight, keepaspectratio]
    {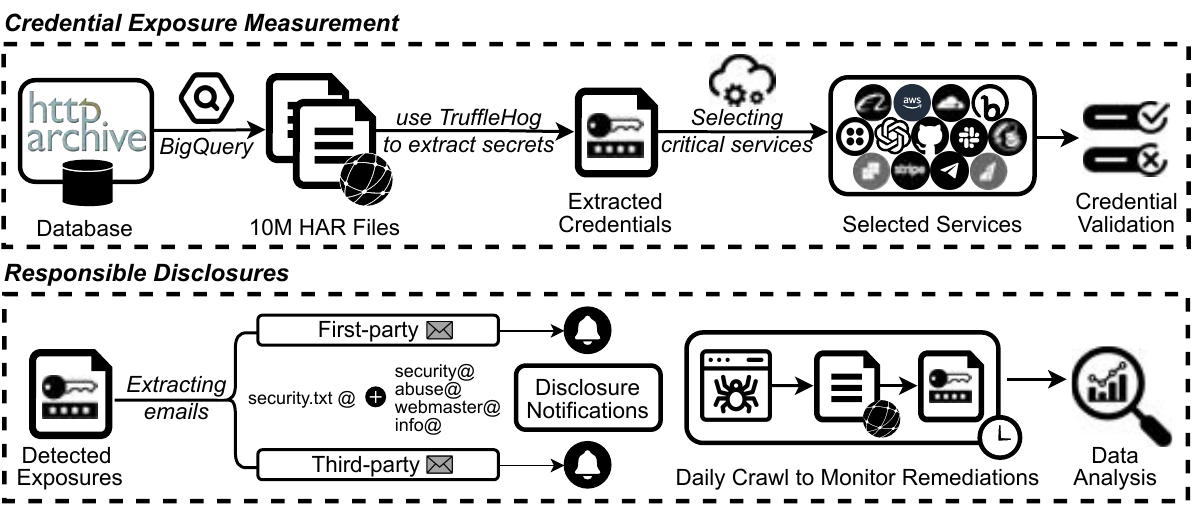}
    \vspace{-5mm}
    \caption{Measurement and Disclosure Pipeline}
    \label{fig:methodology}
\end{figure}

\subsection{Identifying Exposed Credentials}
\label{sec:identifying-exposed-credentials}

We begin by outlining our methodology for identifying exposed credentials and verifying their validity (\Cref{fig:methodology}).

\fakeparagraph{Data Collection} \quad
To measure whether credentials are exposed on the web, we analyze the HTTP Archive dataset, which collects full page loads of websites from the Chrome User Experience Report (CrUX)~\cite{CrUX2025} (which captures the most visited websites by Chrome users~\cite{ruth2022toppling}). The dataset includes network requests, responses, cookies, and headers for all loaded resources, including JavaScript files, stylesheets, and other assets. The crawl data is available in the \textit{HAR} (HTTP Archive) format: a JSON representation that includes the raw request and response data for all resources fetched during each crawl. 
By capturing this full execution, it enables dynamic analysis of the fully rendered webpage.
For this study, we use the September 2025 dataset. Each crawl in HTTP Archive typically contains two pages per origin: the landing page and an inner page (e.g., \textit{http://www.foo.com/bar.html}). Due to its significant size, overhead and higher redundancy, we restrict our analysis to the landing pages only. After filtering, our dataset consists of 11.9M~hostnames (\textit{subdomain.example.com}) and 9.3M~domains (\textit{example.com}). %

\fakeparagraph{Identifying Exposed Credentials} \quad
\label{subsec:identifying-credential-exposure}
Service providers often use recognizable schemes for their credentials (e.g., Stripe's \texttt{sk\_live} prefix), which enable identification through pattern-based techniques like regular expressions. Several tools exist for detecting such exposures. We employ the open-source tool \textit{TruffleHog} (v3.90.8)~\cite{TruffleHog2025}, which can identify over 800~secret types.
We selected \textit{TruffleHog} for its ability to detect and capture not only single-component API keys but also compound credentials. For instance, AWS credentials typically consist of two parts: an \texttt{Access Key ID} (e.g., beginning with \texttt{AKIA}) and a corresponding \texttt{Secret Access Key}. %

\vspace{2pt}
\fakeparagraph{Exposure Localization} \quad
To identify where credentials appear within web resources, we designed an automated \textit{exposure localization pipeline} that processes raw HTTP Archive (HAR) files. In the first stage, the pipeline runs \textit{TruffleHog} in offline mode without validation. \textit{TruffleHog} scans each HAR file and emits detected credential patterns, including those appearing within request URLs, response bodies, JavaScript payloads, or embedded resources. As part of this process, the pipeline attempts to decode common encodings (e.g., URL-encoding and Base64) in request and response fields. 
In a subsequent step, our pipeline processes each detection and traverses the corresponding HAR file to determine the precise location and context of exposure. For every identified credential, we record whether it appears in the request URL, response body, or other components of a webpage. %

\begin{table}[t]
  \centering
  \small
  \caption{Services and Their Verification Endpoints}
  \vspace{-8pt}
  \label{tab:services}
  \begin{tabularx}{\linewidth}{l X}
    \toprule
    \textbf{Service} & \textbf{Verification Endpoint(s)} \\
    \midrule
    Alibaba   & https://ecs.aliyuncs.com (DescribeRegions) \\
    Azure     & https://login.microsoftonline.com/\%s/oauth2/v2.0/token \\
    AWS       & https://sts.amazonaws.com (GetCallerIdentity) \\
    Bitly     & https://api-ssl.bitly.com/v4/user \\
    \makecell[l]{Cloudflare\\} & \makecell[l]{https://api.cloudflare.com/client/v4/user/tokens/verify\\https://api.cloudflare.com/client/v4/user} \\
    GitHub    & https://api.github.com/user \\
    Mailchimp & https://\{dc\}.api.mailchimp.com/3.0/ \\
    OpenAI    & https://api.openai.com/v1/me \\
    RazorPay  & https://api.razorpay.com/v1/items?count=1 \\
    SendGrid  & https://api.sendgrid.com/v3/scopes \\
    Slack     & https://slack.com/api/auth.test \\
    Stripe    & https://api.stripe.com/v1/charges \\
    Telegram  & https://api.telegram.org/bot\{TOKEN\}/getMe \\
    Twilio    & https://verify.twilio.com/v2/Services \\
    \bottomrule
  \end{tabularx}
  \vspace{-5pt}
\end{table}

\vspace{2pt}
\fakeparagraph{Exposure Validation} \quad
After the initial offline detection phase, our analysis revealed a diverse range of credential types embedded across web resources. From these, we selected a subset of high-impact secrets for further validation (Table~\ref{tab:services}). %
We prioritized providers whose keys grant (1) control over cloud infrastructure (Alibaba, AWS, Azure, and Cloudflare), (2) direct financial or transactional access (RazorPay and Stripe), (3) integration with development or supply-chain systems (Bitly and GitHub), and (4) access to communication, data-delivery, and AI platforms that could be leveraged for large-scale abuse or privacy violations (Mailchimp, OpenAI, SendGrid, Slack, Telegram, and Twilio). 

To verify credential validity, our pipeline builds on \textit{TruffleHog}'s built-in verification functionality, which checks whether a credential is active using predefined, non-destructive API calls. These checks require authentication but do not expose, modify, or log sensitive data. For example, when validating Amazon Web Services (AWS) credentials, the pipeline calls the \textit{GetCallerIdentity} endpoint, which simply returns the account identifier associated with the key. The validation relies solely on the HTTP status code to determine authenticity, without inspecting or storing response data.

Before enabling validation, we manually reviewed the corresponding validator implementations within \textit{TruffleHog} to ensure that all API interactions were safe, non-destructive, and compliant with responsible disclosure practices~\cite{owasp_disclosure}. During validation, the pipeline re-runs \textit{TruffleHog} in verification mode, activating only the relevant detectors. A credential is marked as valid once the corresponding service confirms its authenticity through a successful verification response (e.g., status code {\tt 200}).
This confirmation step determines the verified count we report throughout.

\vspace{2pt}
\fakeparagraph{Data Enrichment} \quad
We use several additional datasets to enrich our analysis: (1) HTTP Archive to identify the technologies utilized on the webpages, which relies on Wappalyzer~\cite{wappalyzer_ha} to detect both client- and server-side software; (2) Cisco Umbrella’s~\cite{cisco_umbrella} domain categorization service to classify webpages based on their content and security-related categories; (3) Tranco~\cite{LePochat.Tranco.2019} toplist to determine the popularity rank of the analyzed webpages; (4) To infer the country association of each domain, we use the CrUX dataset~\cite{CrUX2025}.

\vspace{2pt}
\fakeparagraph{Limitations}\quad
In our study, we analyzed keys for accessing only 14~services where we could confirm validity despite identifying 557~credential types. As a result, all findings presented in this paper constitute an explicit lower bound on the true prevalence of exposed credentials, as we cover only 14 service types and count only credentials active at validation time (excluding already-revoked ones). Websites likely further contain additional credentials outside of this scope. Furthermore, our detection relies on TruffleHog, which may not capture all credentials, e.g., credentials stored in encoded or encrypted formats, or credentials delivered in fragments and reassembled at runtime. Despite extensive efforts, our responsible disclosure process may not have reached all affected organizations. In addition, we limited disclosures to verified credentials, as both our results in \Cref{sec:results} and prior work~\cite{basak2023comparative} highlight the inherent false-positive rates in pattern-based credential detection, and we aimed to avoid issuing false-positive notifications to organizations. The actual security impact is likely considerably higher than captured in this conservative study.

\subsection{Longitudinal Analysis}
\label{sec:methodology-longitudinal}
To assess the persistence of exposed credentials over time, we analyze historical data from HTTP Archive's BigQuery dataset. For each validated credential, we extract the corresponding website and a truncated substring of the secret (the first 15 characters, such as \texttt{AKIASECRET12345} for AWS) to avoid handling full sensitive values while maintaining sufficient entropy for reliable matching. Using this substring, we locate and re-visit the same webpages in historical HTTP Archive crawls up to five years prior to our September 2025 dataset. This enables us to identify instances where the same credentials appeared earlier (indicating prolonged exposure).

\subsection{Responsible Disclosure}
\label{sec:methodology-responsible-disclosures}

We disclose identified credentials to affected parties and measure their remediation. We describe our approach below (\Cref{fig:methodology}):

\vspace{2pt}
\fakeparagraph{Identifying Disclosure Recipients} \quad
We leverage email for responsible disclosure, consistent with past studies~\cite{el2025file, stock2018notification, cetin2016understanding, cetin2017make, durumeric2014matter}.
To identify appropriate recipient addresses, we use (1) \texttt{security.txt} when available and (2) generic email addresses from prior disclosure research~\cite{stock2018notification, stock2016hey}.
The \texttt{security.txt} standard~\cite{securitytxt} (RFC 9116~\cite{rfc9116}) allows websites to provide machine-parsable files describing vulnerability disclosure practices, making it easier for researchers to report vulnerabilities. We identify the \texttt{security.txt} file of a website by looking at the root of a website (i.e., \texttt{/security.txt}) as well as at the well-known path. %
We also notify the vulnerable websites via security- and administrator-oriented addresses (\texttt{security@}, \texttt{abuse@}, and \texttt{webmaster@}) per RFC 2142~\cite{rfc2142}.
Additionally, we include \texttt{info@} to reach out to the organization's front office.
Contacting these addresses ensures our notification reaches the relevant administrator who can understand the security risk and facilitate timely mitigation.
In total, we notified 2,435~vulnerable entities (2,279 first-party webpages and 156 third parties) using 9,746 email addresses.

\vspace{2pt}
\fakeparagraph{Content of Disclosure Email} \quad
Figure~\ref{fig:email-template} in the Appendix shows a sample disclosure template.
As proof of exposure, we include service name, partial credential prefix, associated URL, and the observed webpage for each credential.
Partial credentials prevent further exposure if the email is accessed by someone unauthorized.
We suggest two remediation approaches: (1) removal and (2) revocation.
We also request feedback regarding (1) the cause of exposure and (2) awareness of the same.
After one week, we send a reminder.

\vspace{2pt}
\fakeparagraph{Monitoring Remediation} \quad
To understand the effect of our notifications and organizations' mitigation, we monitor remediation after disclosure~\cite{li2016you, stock2016hey, zeng2019fixing, ccetin2019tell, maass2021effective}.
We build a crawler (Appendix~\ref{app:craler-details}) that visits the websites with exposures and collects the network traffic data (i.e., network requests, responses, redirects, headers, and payloads), and re-scans the collected data to identify any prevailing credential exposures (Section~\ref{subsec:identifying-credential-exposure}). 
Additionally, we also perform a day-to-day verification of the exposed credentials detected at the beginning to measure revocation of exposed credentials.
This webpage scanning and the subsequent credential verification are repeated daily, starting from the day we send our first disclosure email (Day 0), allowing us to analyze the organization's approach to addressing the reported exposure. 
We perform daily monitoring for the first two weeks and then perform an additional scan after 6 months (Day 172) to understand the temporal behavior.

\section{Results}
\label{sec:results}

\begin{table}[t]
\centering
\small
\footnotesize
\setlength{\tabcolsep}{4pt}
\caption{Verified Credential Exposures by Service}
\label{tab:validated-secrets}
\vspace{-3mm}
\begin{tabular}{@{}lrrrrr@{}}
\toprule
\textbf{Service} & \textbf{ Identified} & \textbf{ Verified} & \textbf{Verified \%} & \textbf{ Sites} & \textbf{Top Site Rank} \\
\midrule
AWS        & 1,515  & \textbf{283} & 19\%       & 4,693 & Top 500 \\
Azure      & 377    & \textbf{73}  & 19\%       & 92    & Top 500 \\
Alibaba    & 569    & \textbf{29}  & 5\%        & 42    & Top 10K \\
Bitly      & 207    & \textbf{57}  & 28\%       & 400   & Top 5K \\
Cloudflare & 45,046 & \textbf{20}  & $<$0.1\%   & 28    & Top 50K \\
GitHub     & 1.9M   & \textbf{119} & $<$0.1\%   & 272   & Top 500 \\
Mailchimp  & 492    & \textbf{124} & 25\%       & 142   & Top 10K \\
OpenAI     & 1,037  & \textbf{181} & 18\%       & 256   & Top 5K \\
RazorPay   & 1,768  & \textbf{134} & 8\%        & 166   & Top 10K \\
SendGrid   & 255    & \textbf{100} & 40\%       & 2,898 & Top 500 \\
Slack      & 130    & \textbf{66}  & 51\%       & 120   & Top 5K \\
Stripe     & 654    & \textbf{277} & 42\%       & 322   & Top 1K \\
Telegram   & 219    & \textbf{186} & 85\%       & 257   & Top 500 \\
Twilio     & 295    & \textbf{99}  & 34\%       & 116   & Top 100K \\
\midrule
\textbf{Total} & \textbf{2M} & \textbf{1,748} & \textbf{0.09\%} & \textbf{9,804} & -- \\
\bottomrule
\end{tabular}
\vspace{-8pt}
\end{table}

In this section, we present our results on identified credentials.
Overall, we find approximately 2M candidate credentials across the analyzed services, of which 1,748 (0.09\%) are confirmed valid through provider APIs (Table~\ref{tab:validated-secrets}).
We examine this discrepancy and its underlying causes in the following section.

\vspace{2pt}
\fakeparagraph{Potential Credentials}\quad For some services, there is a large discrepancy between TruffleHog-identified credentials and verified credentials, which reflects the trade-off of detection methods that emphasize coverage over precision. GitHub dominates the discarded pool (roughly 98\% of candidates), where its broad 40-character hexadecimal pattern matches non-credential identifiers pervasive on the web such as session and tracking IDs, embedded digests, ETags, and asset fingerprints; Cloudflare, Bitly, and Alibaba over-match similarly. As a result, GitHub tokens exhibit a true-positive rate of only 0.0001\% and account for 1.9M candidate credentials.

For services with distinctive formats such as Stripe, OpenAI, AWS, and SendGrid, discarded candidates are well-formed keys that appear \textit{syntactically} correct but had already been revoked or rotated before our validation.
Narrowly specified patterns verify far higher (Stripe 42\%, Telegram~85\%).
This is largely driven by the fact that some services like Telegram use token formats that are unambiguous.
Telegram's pattern follows a fixed structure consisting of a numeric identifier, a colon, and an alphanumeric key (e.g., \texttt{123456789:ABCDefGhIjKlm}) which allows the detector to identify tokens with a higher precision. Thus, our exploratory analysis indicates that the high false-positive rate stems not only from \textit{TruffleHog}’s generic detection patterns but also from the heterogeneity of credential formats themselves.

\subsection{Verified Credentials}
Next, we focus on verified credentials that represent genuine security exposures.
In total, we identify 1,748~verified credentials across 9,804~websites and 5,693~domains.
Exposure is widespread across service categories, with cloud services (e.g., AWS, Cloudflare) and payment services (e.g., Stripe, RazorPay) accounting for the majority of credentials.
AWS credentials represent more than 16\% of all verified exposures on 4,693~websites.
Email and communication services such as SendGrid and Twilio also appear frequently, with a significant portion of their exposures originating from embedded third-party resources. Credential exposures occur not only within first-party code but also within external dependencies, underscoring that risks extend beyond local development practices.
Several verified credentials were found on highly popular domains (e.g., within the Tranco Top~\num{500} websites), showing that even well-maintained and widely used web services are not immune to inadvertent credential exposure.
In the remainder of this paper, we focus explicitly on verified credentials.

\subsection{Ecosystem Analysis}
\label{sec:toplist}

In this section, we begin to analyze how credentials become exposed by distinguishing between first-party resources—those directly hosted by websites—and third-party resources that are externally embedded or loaded. 
Exposures associated with first-party resources occur predominantly on less popular websites, whereas those linked to third-party resources are more frequently observed also among higher-ranked websites (\Cref{fig:top-list}). This indicates structural differences in how credentials become exposed:

\begin{figure*}
    \centering
    \includegraphics[width=0.82\linewidth]{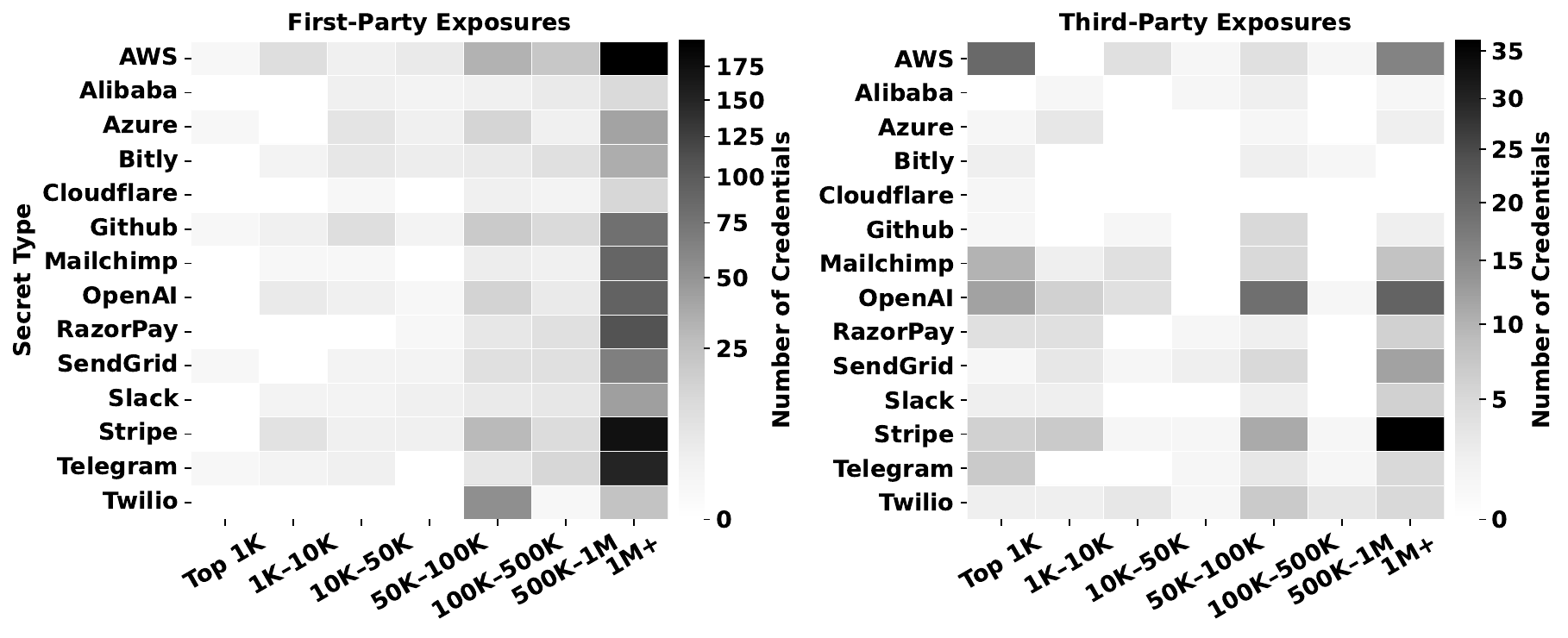}
    \vspace{-3mm}
    \caption{Credential Exposures By Website Popularity---\textnormal{First-party exposures are common at all ranks. Third-party exposures are frequent, particularly on popular third-party integrations, and often result from first-party misconfigurations.}}
    \label{fig:top-list}
\end{figure*}

\fakeparagraph{Exposures from First-Party Resources} \quad
Next, we consider exposures across first-party and third-party resources. We find that \res[a2.2]{84\%} of credentials are present in first-party resources.
Surprisingly, we find \res[a2.3]{478} (5\%) cases where credentials appear directly in the initial HTML response. Our findings show notable differences across services in where credentials are exposed, whether through first-party or third-party resources. The majority of distinct credentials for GitHub (97\%), Cloudflare (\res[e2.4]{95\%}), and Bitly (\res[e2.4]{95\%}) appear within first-party resources, suggesting that these credentials are mostly embedded directly into webpage's code or configuration. As shown in~\Cref{fig:top-list}, the number of credential exposures increases as website popularity declines. At the same time, we also identify first-party exposures among highly ranked websites, indicating that credential management issues persist even among popular domains.

\fakeparagraph{Exposures from Third-Party Resources} \quad
Our results show 16\% of exposed credentials stem from third-party resources. OpenAI credentials exhibit the highest share of third-party exposures, with \res[e2.4]{24\%} of distinct OpenAI credentials appearing in externally loaded resources. Similarly, Twilio (24\%), Mailchimp (23\%), Alibaba (23\%), Stripe (22\%), and SendGrid (22\%) also show notable third-party exposure rates. The majority of these services support user-facing features that might be intended for client-side use, and their appearance in web resources may therefore be intentional, even if unsafe. We further address this phenomenon in our disclosure analysis (\Cref{subsec:disclosures-root-cause-analysis}), drawing on feedback from affected parties. 

Considering which third-party origins are most frequently associated with exposures, we find that \textit{likr.tw} accounts for the largest number of exposed credentials with \res[a2.5]{3,752} cases, followed by \textit{bestfreecdn.com} (\res[a2.5]{\num{2633}}), \textit{googletagmanager.com} (\res[a2.5]{283}), and \textit{cloudfront.net} (\res[a2.5]{137}). 
 When considering distinct credentials, the third-party domains {\textit{googleapis.com} (21~credentials), \textit{googletagmanager.com} (20), \textit{amazonaws.com} (11), and \textit{cloudfront.net} (11)} appear most frequently. 
 
This discrepancy reflects differences in the responsibility for exposure. In the case of \textit{likr.tw}, a single credential is reused across multiple websites through a shared script, meaning that the third party is directly responsible for the widespread exposure. In contrast, domains hosting more distinct credentials such as \textit{googletagmanager.com} are \textit{unlikely} to be the original source of exposure; rather, they reflect first-party misconfigurations where credentials are embedded within assets deployed. Our results indicate that while third-party domains often act as secondary distribution points for credentials originating from site-managed resources, they still play a substantial role in the overall exposure landscape.

\subsection{Exposure Environment}
\label{sec:exp-environment}

Next, we analyze the technical environment to identify the root causes of these exposures.

\fakeparagraph{File Types}  \quad
We next examine the types of files in which credentials are exposed to better understand their technical context. The majority of exposures occur within JavaScript resources (\res[e3.1]{84\%}), followed by HTML (\res[e3.1]{8\%}) and JSON (\res[e3.1]{7\%}) files.
Interestingly, we identify isolated credentials in unconventional locations (e.g., a verified GitHub access token embedded in a CSS file). We find that a large share of credentials are embedded within bundled JavaScript files. Bundling is a common optimization technique that combines multiple scripts into a single resource to improve performance~\cite{rackwebpack}. To examine this systematically, we identify bundled files using runtime fingerprints inserted by popular build tools, based on the fingerprint list provided by prior work~\cite{rackwebpack}. For example, bundles produced by Webpack contain the marker \texttt{\_\_webpack\_require\_\_}.
Our analysis shows that \res[a4.1]{62\%} of credential exposures in JavaScript files occur within bundles created by popular build tools such as Webpack. Note that the actual results are likely higher, as we do not cover all custom bundlers.

These findings underscore the necessity of dynamic analysis. Since most credential exposures are introduced dynamically---such as during the build and deployment process, materializing within these compiled bundles---the traditional methodology of scanning static source code repositories used by prior work would miss them.

\res[e3.4*]{}
\begin{table*}[t]
  \centering
  \small
  \caption{Top Technologies and Software Categories for All Web (HTTP Archive) vs.\ Sites with Exposed Credentials}
  \vspace{-10pt}
  \label{tab:ha-vs-leaks-top}
  \begin{tabularx}{\linewidth}{>{\raggedright\arraybackslash}X r >{\raggedright\arraybackslash}X r @{\hspace{1.5em}} >{\raggedright\arraybackslash}X r >{\raggedright\arraybackslash}X r}
    \toprule
    \multicolumn{4}{c}{\textbf{Technologies}} & \multicolumn{4}{c}{\textbf{Software Categories}} \\
    \cmidrule(lr){1-4} \cmidrule(lr){5-8}
    \textbf{HA (name)} & \textbf{HA (\%)} & \textbf{Exposures (name)} & \textbf{Exp. (\%)} & 
    \textbf{HA (name)} & \textbf{HA (\%)} & \textbf{Exposures (name)} & \textbf{Exp. (\%)} \\
    \midrule
    jQuery           & 71\% & Google Analytics     & 70\% & JS Libraries   & 88\% & Miscellaneous          & 88\% \\
    Open Graph       & 66\% & Open Graph           & 70\% & Miscellaneous          & 85\% & JS Libraries   & 83\% \\
    Google Analytics & 53\% & core-js              & 69\% & Analytics              & 63\% & JS Frameworks  & 76\% \\
    PHP              & 53\% & jQuery               & 65\% & Programming Lang.  & 63\% & Analytics              & 74\% \\
    Google Font API  & 44\% & Cloudflare           & 59\% & Web Servers            & 59\% & CDN                    & 69\% \\
    \bottomrule
  \end{tabularx}
  \vspace{-3mm}
\end{table*}

\fakeparagraph{Software} \quad
Next, we analyze server- and client-side software components. We find an average of \res[e3.5a]{17} software components per website in the broader HTTP Archive dataset, compared to \res[e3.5b]{22} on websites exposing credentials (interquartile ranges \res[e3.5c]{14–24} vs.~\res[e3.5d]{17–27}).
This indicates that websites with exposed credentials tend to employ a broader mix of technologies, reflecting higher integration density and aligning with observations that greater software diversity often correlates with increased configuration risk~\cite{demir_our_2021}.
We also observe differences in the types of software between the two groups (\Cref{tab:ha-vs-leaks-top}).
Websites with exposed credentials show a stronger reliance on JavaScript frameworks and client-side libraries, whereas widely used foundational technologies dominate the web. 
These findings are consistent with our earlier observation that a large portion of exposures originate within the JavaScript environment.

\fakeparagraph{Credential Encoding} \quad
Next, we examine how credentials are represented within web resources to assess their encoding or representation form. Our framework supports three encoding types for a given credential: \textsc{Plain}, \textsc{Escaped\_Unicode} (e.g., \texttt{\textbackslash u006b\allowbreak\textbackslash u0065\allowbreak\textbackslash u0079}
for \texttt{key}), and \textsc{Base64} (e.g., \texttt{a2V5}). Most exposed credentials appear without substantial encoding, with 67\% detected as \textsc{Plain}, 18\% as \textsc{Escaped\_Unicode}, and 15\% as \textsc{Base64}. In practice, this means that the majority of exposed credentials are human-readable, while the rest offer only cosmetic protection and can be easily reversed, and provide no real protection.
 
\fakeparagraph{Geographic Distribution} \quad Next, we analyze the geographical distribution of distinct exposed credentials to assess how exposure varies across countries using the CrUX dataset~\cite{CrUX2025}. 
\Cref{tab:country-share-two-top5} compares each country’s share of all web domains in the HTTP Archive dataset with its share of domains containing exposed credentials in our measurements.
While the overall distribution aligns with global web share, several countries show a markedly higher representation among domains with exposures. 
The United States, for instance, contributes 20\% of all exposed domains compared to 9\% of globally active domains, and India accounts for 16\% of exposed domains versus 5\% of global domains.

\res[a4.1+2]{}
\begin{table}[t]
\centering
\small
\caption{Distribution of Domains by Country}
\label{tab:country-share-two-top5}
\vspace{-3mm}
\begin{tabular}{lrlr}
\toprule
\multicolumn{2}{c}{\textbf{Global Domain Share}} &
\multicolumn{2}{c}{\textbf{Vulnerable Domains}} \\
\cmidrule(lr){1-2} \cmidrule(lr){3-4}
\textbf{Country} & \textbf{Share (\%)} & \textbf{Country} & \textbf{Share (\%)} \\
\midrule
United States & 9.1 & United States & 20.0 \\
India & 4.7 & India & 15.7 \\
Indonesia & 4.4 & Italy & 5.6 \\
Brazil & 3.5 & Brazil & 4.3 \\
Germany & 3.5 & United Kingdom & 3.8 \\
\bottomrule
\vspace{-8mm}
\end{tabular}
\end{table}

\begin{figure}
    \centering
    \includegraphics[width=1\linewidth]{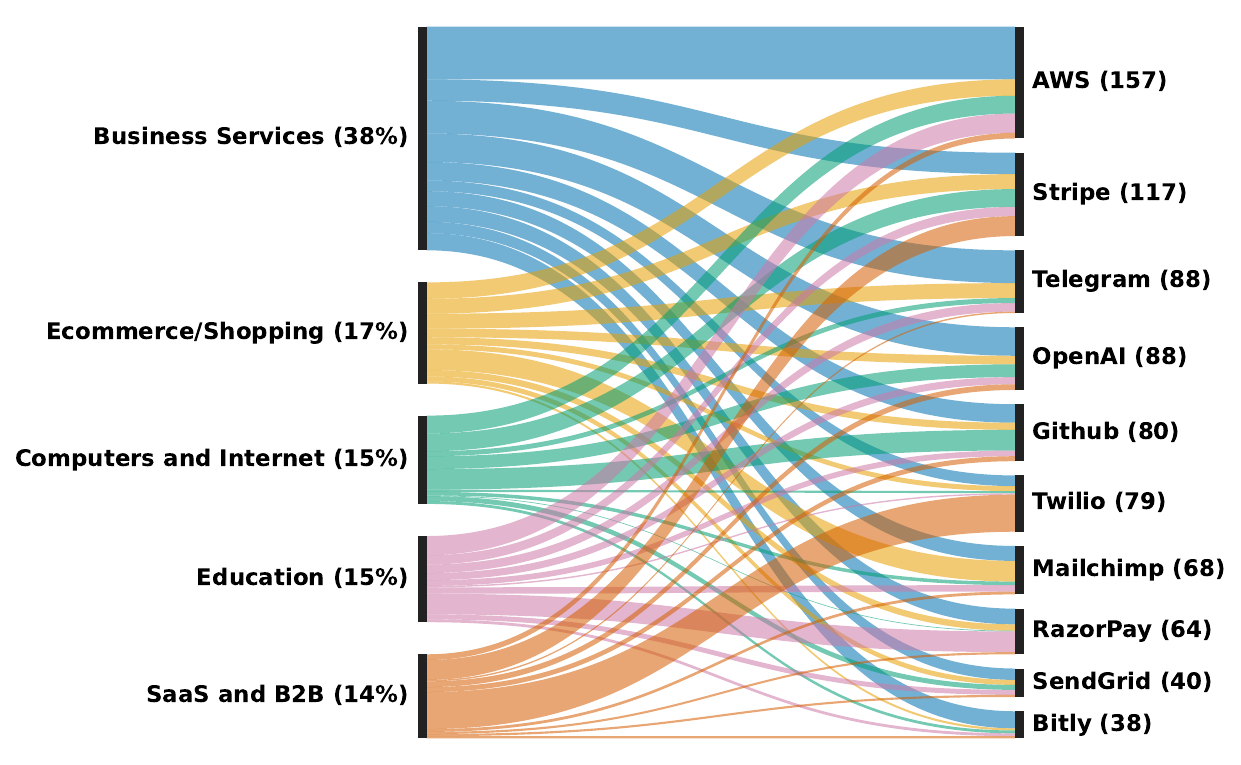}
    \vspace{-6mm}
    \caption{Credential Exposure by Category and Service}
    \label{fig:category-service-sankey}
\end{figure}

\fakeparagraph{Types of Websites}\quad
To contextualize the affected domains, we classify them by their primary category. The five most common website categories among exposed domains are Business Services (17\%), E-commerce/Shopping (8\%), Computers and Internet (7\%), Education (7\%), and SaaS and B2B (6\%).
Figure~\ref{fig:category-service-sankey}  illustrates the relationship between these website categories and the most frequently exposed credential types.
The distribution indicates that each category relies on a distinct combination of external providers.
For instance, business-oriented websites show strong associations with infrastructure providers such as AWS and service platforms like OpenAI. 
Domains in the Computers and Internet category are most commonly linked to GitHub.
SaaS and B2B platforms, in turn, maintain close ties with communication and payment providers such as Twilio and Stripe, consistent with their use of third-party APIs for customer interaction and transaction management.
Thus credential exposures are not uniform but follow the specific dependency patterns of each domain category.

\begin{figure}[t]
  \centering
  \includegraphics[width=0.98\columnwidth]{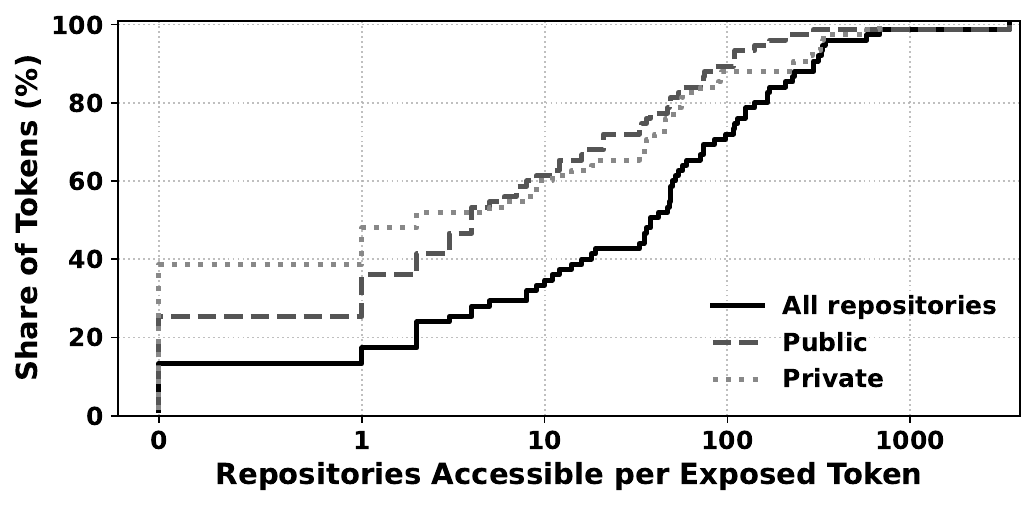}
    \vspace{-3mm}
  \caption{Blast Radius per Exposed GitHub Token}
  \label{fig:blast-radius-cdf}
\end{figure}

\subsection{Quantifying the Blast Radius}
\label{sec:blast-radius}
Next, we analyze the reach of a single exposed credential by resolving each GitHub token to the repositories it grants access to. The 119 GitHub credentials we identified in this work provide access to 9,720~repositories (4,270 private and 5,450 public) across 241~distinct accounts, with admin access to 7,942 and push access to all of them. Beyond source code, the tokens also permit modifying CI/CD workflows (20 tokens), publishing packages (20), deleting repositories (9), and administering entire organizations (13) or enterprises (8). 
This reach is highly skewed across tokens, ranging from none to 3,496~repositories on a single token (Figure~\ref{fig:blast-radius-cdf}).
The reachable repositories include public projects with more than 15k~stars, and in one case a single token grants admin access to a company's complete infrastructure-as-code and its payment services.

\subsection{Case Studies}
\label{sec:case-studies}
To illustrate the real-world impact of exposed credentials, we more deeply consider several specific cases where credential exposure creates systemic risks for financial and critical infrastructure, the software supply chain, and public safety.

\fakeparagraph{Systemic Risk in Financial Infrastructure} \quad
While we identified direct exposures within several international banks, a more critical risk was found in a core infrastructure provider servicing a national banking network (GitHub credentials found in a public JavaScript file). These credentials granted full access to the software stacks used by the major national banks of that country, illustrating a precarious dependency where a single upstream provider's security effectively underpins the integrity of a nation’s entire financial infrastructure.
The vendor did not respond to our disclosure but revoked the credentials.

\fakeparagraph{Cyber-Physical and Hardware Supply Chains} \quad
Next, we illustrate how exposed credentials on the web can lead to concrete supply-chain risks. We identified GitHub credentials belonging to a firmware developer responsible for a firmware for a radio-transmitter which is used by numerous manufacturers of drones, toys, and other remote-controlled devices globally. The exposed credentials granted full repository privileges for firmware distribution. An adversary obtaining these credentials could modify source code or firmware binaries, inject malicious functionality, or tamper with release artifacts distributed to downstream manufacturers. Such a compromise would propagate through legitimate update channels and affect physical devices globally, effectively transforming a single credential exposure into a large-scale supply-chain threat.
We contacted the developer, who confirmed awareness of the issue and initially stated that the credentials were limited to read access for public repositories. However, upon verification, they acknowledged that the credentials had full access. 
The developer promptly revoked the exposed credentials, reviewed the commit history, and confirmed finding no signs of malicious activity.

\fakeparagraph{Global Systemic Finance} \quad
We discovered AWS credentials belonging to a ``Global Systemically Important Bank'' (G-SIB). Such institutions are designated by financial regulators as critical to the stability of the global financial system, as their disruption could propagate risks across markets and other interconnected entities~\cite{BIS_GSIB_2025}.
The exposed credentials granted root access to multiple AWS services, including Lambda, Relational Database Service, Secrets Manager, and Key Management Service. 
A manual investigation of the affected public JavaScript file revealed additional exposed credentials and sensitive data, such as RSA private keys, OAuth2 credentials, and internal service URLs. 
Following our disclosure, the bank acknowledged the issue and promptly removed the credentials from the exposed resources.

\fakeparagraph{National Energy Utilities} \quad
Finally, we identified Azure credentials for a national energy provider (exposed in the landing page's HTML) on an administrative portal and an Access Control Manager. This finding is significant as it reveals an unintended bridge between public-facing web assets and the internal management plane of a national utility. By exposing credentials on the tools designed to govern them, the provider inadvertently turned its security surface into a high-leverage entry point to the digital infrastructure of one of the country's largest utility systems.
The provider did not respond to our disclosure but revoked the exposed credentials.

\fakeparagraph{Public Safety and Privacy-Sensitive Systems} \quad
Beyond financial and infrastructure providers, we found exposures in public-sector systems that process highly sensitive personal data. In one case, we identified Azure credentials for a statewide school safety and threat-management platform (holding student behavioral threat assessments and school-incident records on minors). In a second, we found Azure credentials for a law-enforcement and municipal-court platform (records management and court case management). Both were exposed in publicly accessible JavaScript files. In both, the exposed credential bridged a public-facing web asset to the management plane of a system holding sensitive personal and, in the latter, law-enforcement data---turning a client-side leak into a privacy and public-safety exposure. Neither organization responded to our disclosure, but both revoked the exposed credentials.

\subsection{Exposure Persistence}
\label{sec:results-persistence}

\begin{figure}[htbp]
    \centering
    \includegraphics[width=.96\linewidth]{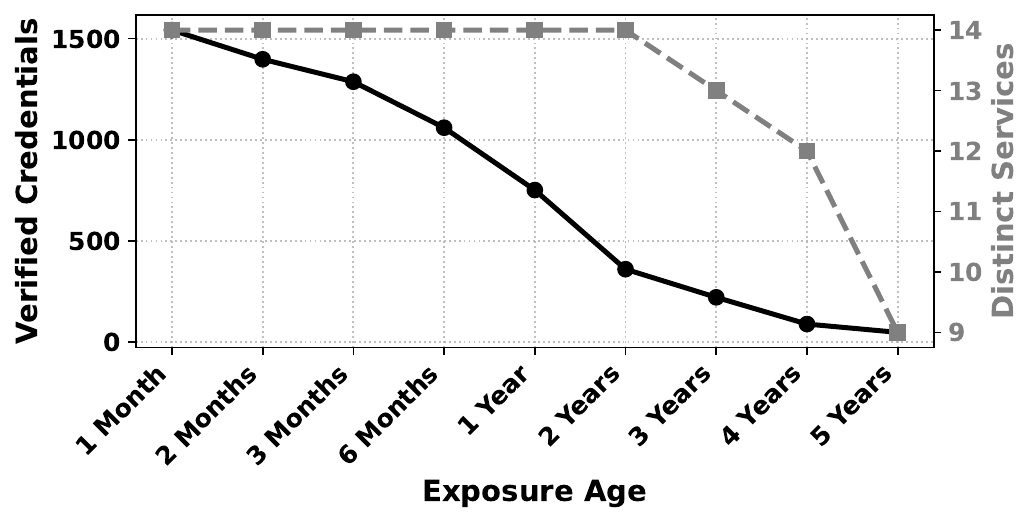}
    \vspace{-3mm}
    \caption{Exposed Credentials and Services Over Time}
    \label{fig:persistence-results}
\end{figure}

Next, we analyze how long verified credentials remain publicly exposed.
To estimate persistence, we examine nine distinct crawl snapshots from HTTP Archive, tracing when each credential first appeared and whether it continued to be accessible over time.
We limit our analysis to nine measurement points due to the financial overhead of large-scale historical queries, which required processing more than 1.5\,PB of data. As can be seen in \Cref{fig:persistence-results}, exposed credentials remain publicly accessible for an average of 12~months (min: 1; SD: 14; max: 60). In the first month after exposure, we observe a 10\% reduction in the number of active credentials, suggesting that only a small fraction of affected providers detect and remove their exposed credentials promptly.

\begin{figure}[htbp]
    \centering
    \includegraphics[width=0.88\linewidth]{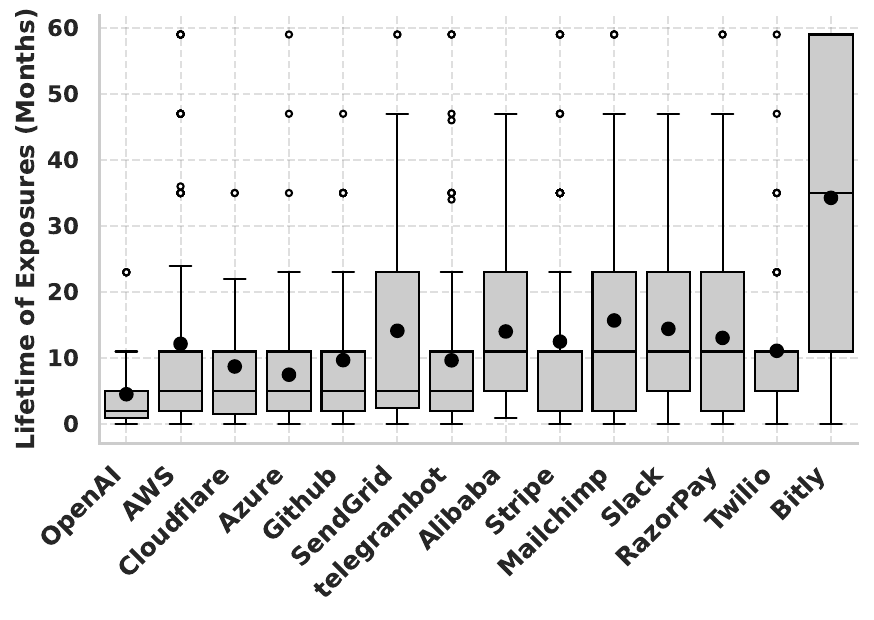}
        \vspace{-4mm}
    \caption{Persistence of Exposed Credentials by Service}
    \label{fig:lifetime}
\end{figure}

A Kruskal–Wallis H-test confirms that exposure lifetimes differ significantly between services ($p<0.001$). Services like Slack and Bitly exhibit longer exposure periods, while cloud credentials (e.g., AWS, Azure) have shorter-lived exposures (\Cref{fig:lifetime}). OpenAI’s API also shows short durations; since its introduction in late 2022~\cite{openai_introducing_chatgpt_2022}, observed lifetimes are limited by the measurement window. Bitly is a notable outlier, persisting significantly longer than other providers. The causes of these differences might stem from several factors. The short observed lifetime does not necessarily imply that websites stop exposing credentials; it may instead indicate revocation or rotation by developers. Some services may deploy automated systems that detect and revoke exposed credentials more quickly than others, and developers may intervene manually once aware of an exposure. 

\section{Responsible Disclosure}
\label{sec:results-disclosure}
 
We disclosed exposed credentials to the affected parties---2,279~first-party and 156~third-party notifications---totaling 9,746~disclosure emails. In this section, we study the effect of our responsible disclosures on overall web security.

\fakeparagraph{Delivery Success} \quad
Out of 9,746~emails sent, we observed a bounce rate of 58\% (5,658~emails), which is lower than previously observed~\cite{cetin2017make, cetin2016understanding, stock2016hey, stock2018notification}.
Across the email variations attempted for each website, we successfully delivered at least one email to 1,668~websites (68.5\% success rate). 
Among the failed 767 websites, we adopted a manual approach for obtaining email addresses associated with the most critical websites belonging to financial institutions, health and medicine, and government and law institutions. We were able to find emails for 61/68 of these websites.
Together, we sent notifications to 71\% of sites (1,729~domains).
As described in our disclosure methodology (Section~\ref{sec:methodology-responsible-disclosures}), exposures on first-parties led to notifications to first-parties, whereas exposures originating from embedded third-party resources led to notifications to the affected third-party, with first parties additionally notified only when they were directly responsible for the integration (e.g., \textit{googletagmanager.com}). This explains the difference between the number of domains we notified and the total number of affected domains reported in~\Cref{tab:validated-secrets}.
Two domains requested to opt out and were removed from our analysis.
In the rest, we focus our analysis by studying the effect of our disclosures on the 1,727 websites in mitigating reported exposures.

\subsection{Impact of Disclosures}

\begin{figure*}[!t]
  \centering
  \subfigure[Effect of Credential Removals]{
      \includegraphics[width=0.48\textwidth]{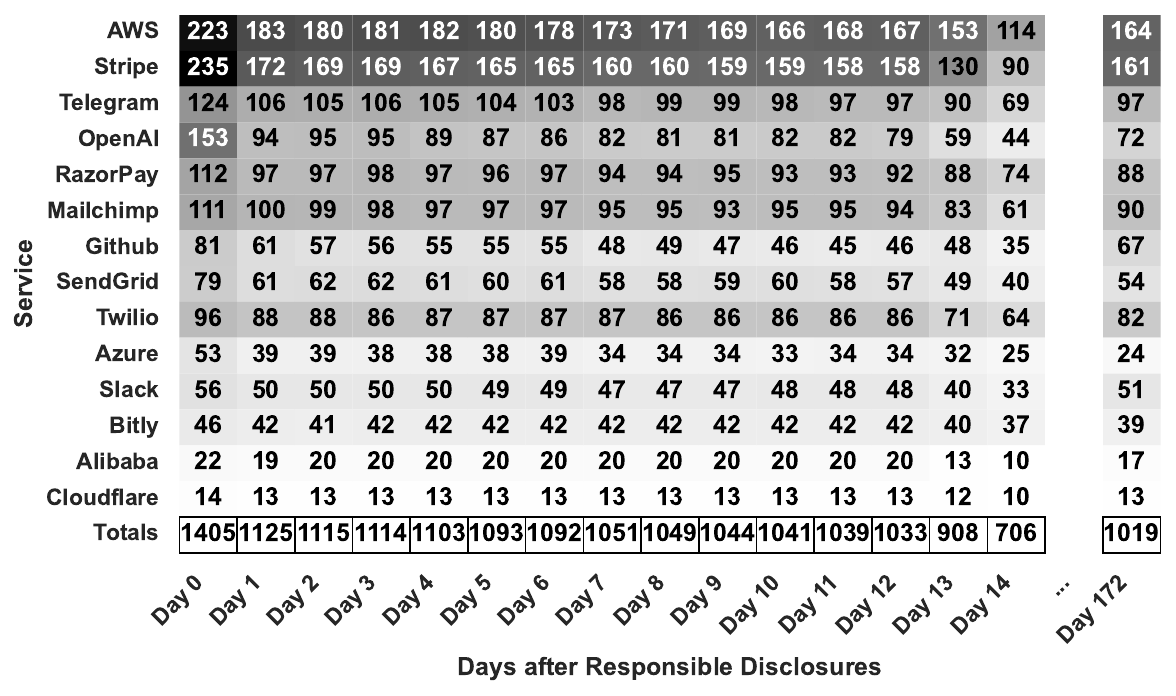}
      \label{fig:a}
  }\hspace{-3mm}%
  \subfigure[Effect of Credential Revocations]{
      \includegraphics[width=0.51\textwidth]{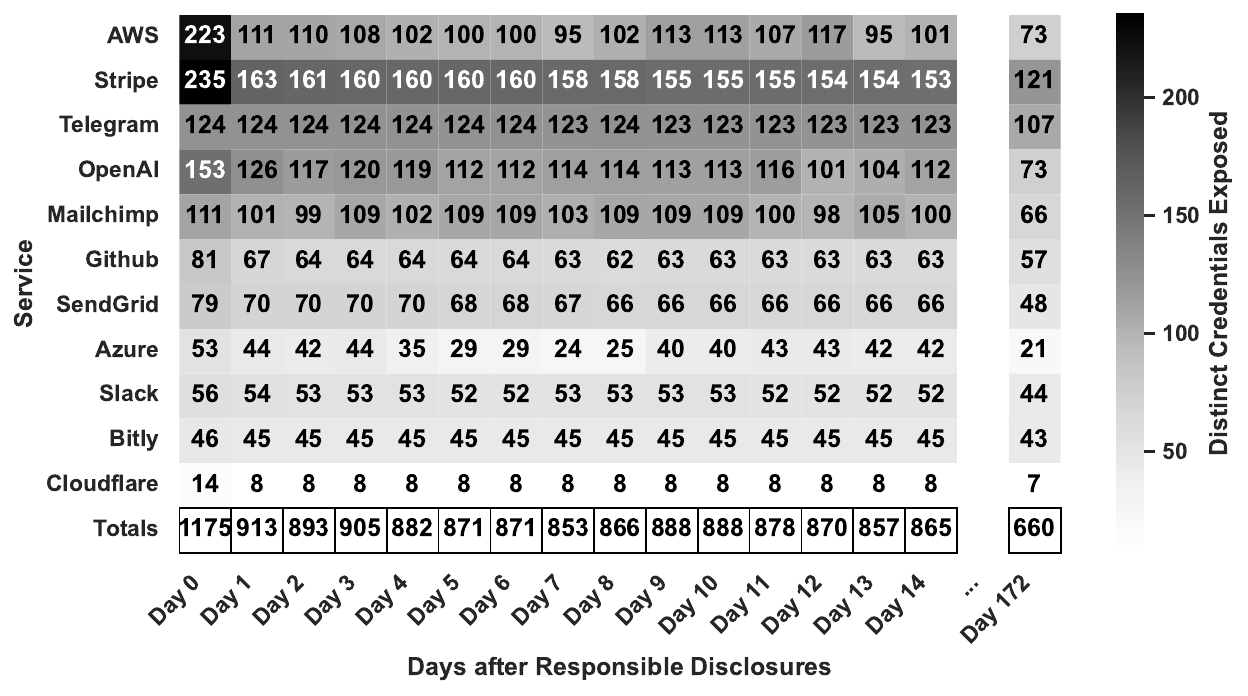}
      \label{fig:b}
  }
  \vspace{-10pt}
   \caption{Remediation Effects of Our Responsible Disclosures---\textnormal{(a) credential removals (b) and credential revocations. (a) depicts total number of exposed credentials found on web each day and (b) shows the daily counts of verified exposed credentials. Day 0 represents the measurement performed before sending the disclosures on the same day.}}
  \label{fig:longitudinal-effect-disclosures}
  \vspace{-3mm}
\end{figure*}

\fakeparagraph{Effect of Disclosures Across Services} \quad
To understand the effect of our disclosures, we study two remediation measures: (1) removal and (2) revocation of the reported credentials.
Removal alone \textit{does not} make a credential safe: cached or archived copies of the webpage persist, leaving an unrevoked credential retrievable. 
We therefore interpret removal without revocation as partial remediation.
We use the data from our daily crawls (Section~\ref{sec:methodology-responsible-disclosures}) to measure the presence or absence of pre-disclosure credentials on affected websites.
Subsequently, we also perform daily verification of these credentials to understand their revocations (Section~\ref{sec:identifying-exposed-credentials}).
Figure~\ref{fig:longitudinal-effect-disclosures} depicts the impact of our disclosures on daily counts of exposed credentials from each service across affected websites in (a) and active credentials in (b). 

\begin{figure}[!bpt]
    \centering
    \includegraphics[width=.97\linewidth]{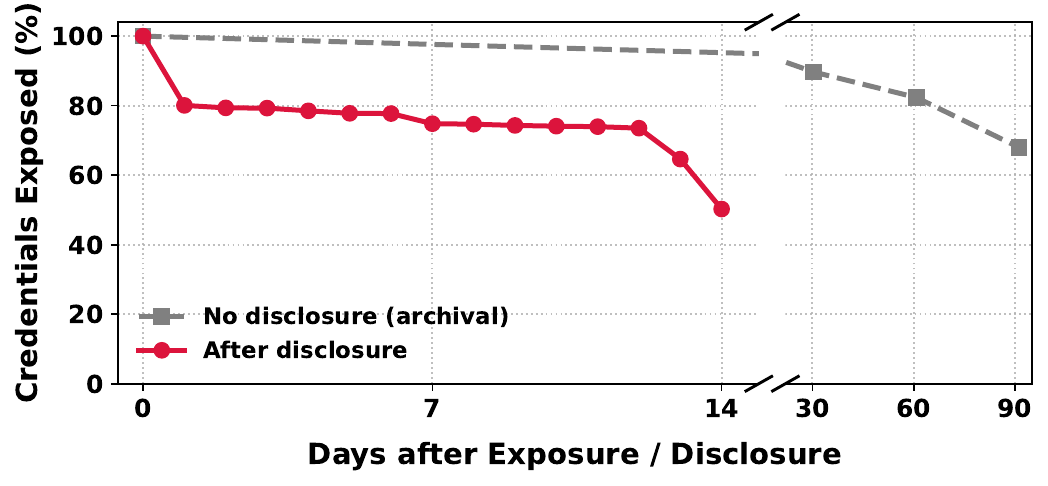}
    \vspace{-3mm}
    \caption{Impact of Responsible Disclosure Over Time}
    \label{fig:disclosure_vs_archive}
\end{figure}

As seen in \Cref{fig:disclosure_vs_archive}, we observe a clear decline in credentials across all services with exposures reducing 50\% from 1,405 (Day 0) to 706 (Day 14), far exceeding the approximately 10\% removed per month without intervention (\Cref{fig:persistence-results}). 
OpenAI shows the largest percentage drop in exposures across services ranging from 153 on Day 1 to 44 by Day 14, a reduction of 71\%, followed by Stripe (62\% removals) and GitHub (57\% removals); while Bitly records the lowest percentage drop at 20\%. %
In terms of absolute count of removed credentials, Stripe credentials are removed the most (145 keys), followed by AWS and OpenAI---with 109 keys each removed by Day 14. 
On average, 47\% of credentials associated with each service were removed. 
On the other hand, credential revocations are less common than the removals with---verified credential exposures dropping from 1,175 to 865---a drop of only about 26\% within 14 days of the disclosure. %

Notably, total non-removed exposures increase from 706 (Day 14) to 1,019 (Day 172) but non-revoked exposure counts show a further decrease from 865 (Day 14) to 660 (Day 172). 
This decline in non-revoked credentials suggests that developers actively revoked some more credentials between Day 14 and Day 172; the increase in non-removed credentials is therefore likely attributable to accidental re-exposures rather than negligence, as developers otherwise demonstrate the remediation efforts. 
It is also important to note that we observe Day 0 totals for removals and revocations to be different, likely because our IP address was blocked by services---Alibaba, RazorPay, and Twilio during credential verification crawl. 
AWS credentials have the highest revocations (122 revocations or 55\%) while Telegram and Bitly credentials are the two least revoked ones, with only 1 revocation each. 
There could be several explanations behind aggregated removals (50\%) to be approximately 2 times the aggregated revocations (26\%)---(1) websites believe credential removals to be sufficient to prevent any further abuse; (2) credentials were no longer in use or were public or read-only; (3) website uses the credentials for more than one purpose, making it cost/resource prohibitive to revoke.

\begin{figure}[t]
    \centering
    \includegraphics[width=0.80\linewidth]{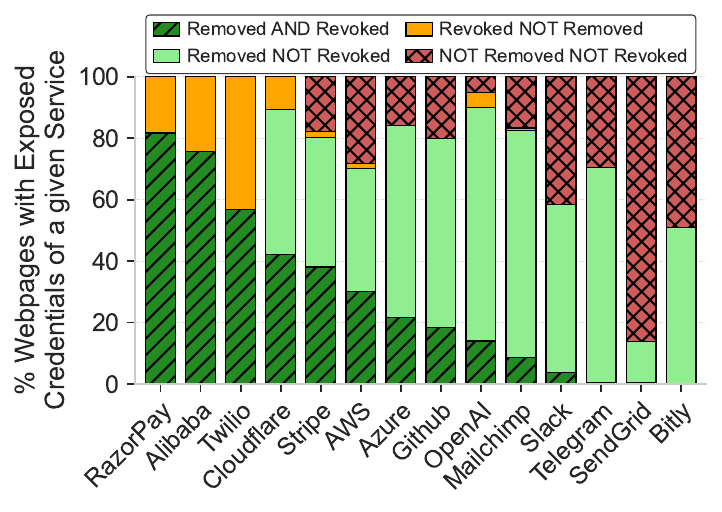}
    \vspace{-7pt}
    \caption{Remediation Response After Disclosure}
    \label{fig:rem-revo-both-none}
    \vspace{-3mm}
\end{figure}

\fakeparagraph{Effect of Disclosures Across Websites} \quad
Next, we analyze the effect of our responsible disclosure across websites.
Figure~\ref{fig:longitudinal-effect-disclosures} represents the aggregated impact of removals and revocations across services, while Figure~\ref{fig:rem-revo-both-none} shows, per service, the fraction of websites that remove, revoke, both, or neither.
Services like GitHub, Open\-AI, Mailchimp, Slack, and Telegram show a larger fraction of credentials being removed but not revoked, suggesting developers often relied on partial fixes that failed to eliminate backend risks. 

Figure~\ref{fig:cdf-website-removals} in the Appendix represents a CDF of remediations across different days post disclosure for all API services.
Credential removals vary significantly by service: nearly 90\% of OpenAI-using websites removed at least one credential, while only 50\% of Bitly- and 15\% of SendGrid-using websites did so, suggesting divergent security responses depending on the provider. 
Except Bitly, we observe an upwards trend in temporal credential removals, showing the overall impact of disclosures.

\fakeparagraph{Geographic Effect of Disclosure Remediations} \quad
To assess geographical variation in remediation, we test the fraction of websites per country that remove and revoke credentials.
Figure~\ref{fig:country-removals-revocations} in the Appendix shows that Italy demonstrates the highest revocations, followed by India and Japan.  
In terms of removals, more than 60\% of South Korean websites removed their exposed credentials, followed by Brazil, Russia, and Taiwan.  
In fact, Russia, Ukraine, and Taiwan are among the bottom three with the lowest revocation rates.

\fakeparagraph{Category and Rank of Websites} \quad
Next, we test how remediation varies with a website's category, comparing binary removal and revocation outcomes across Tranco rank categories and the top-5 website categories using pairwise two-proportion z-tests, with Cohen's h quantifying the effect size.
By rank, credential removals are significantly more common among popular websites (ranked 1K--500K) than among less popular ones (500K--1M and 1M+; $p<0.01$).
Revocations, however, follow a non-monotonic pattern: both the most popular sites (ranked up to top-50K) and the least popular ones (ranked 500K+) revoke significantly less often than sites of average popularity (50K--500K).

This suggests that the most popular sites may integrate the same credential across more complex workflows, making revocation harder, while less popular sites may lack the security maturity or operational capacity to remediate.
By category, Computers, Education, and B2B websites remove exposed credentials significantly more often than Business and E-commerce sites, and E-commerce sites engage in removal or revocation significantly less than other verticals ($p<0.01$).
Taken together, these results indicate that technically oriented verticals exhibit better remediation practices.

\subsection{Root Cause Analysis}
\label{subsec:disclosures-root-cause-analysis}
\begin{figure}
    \centering
    \includegraphics[width=0.9\linewidth]{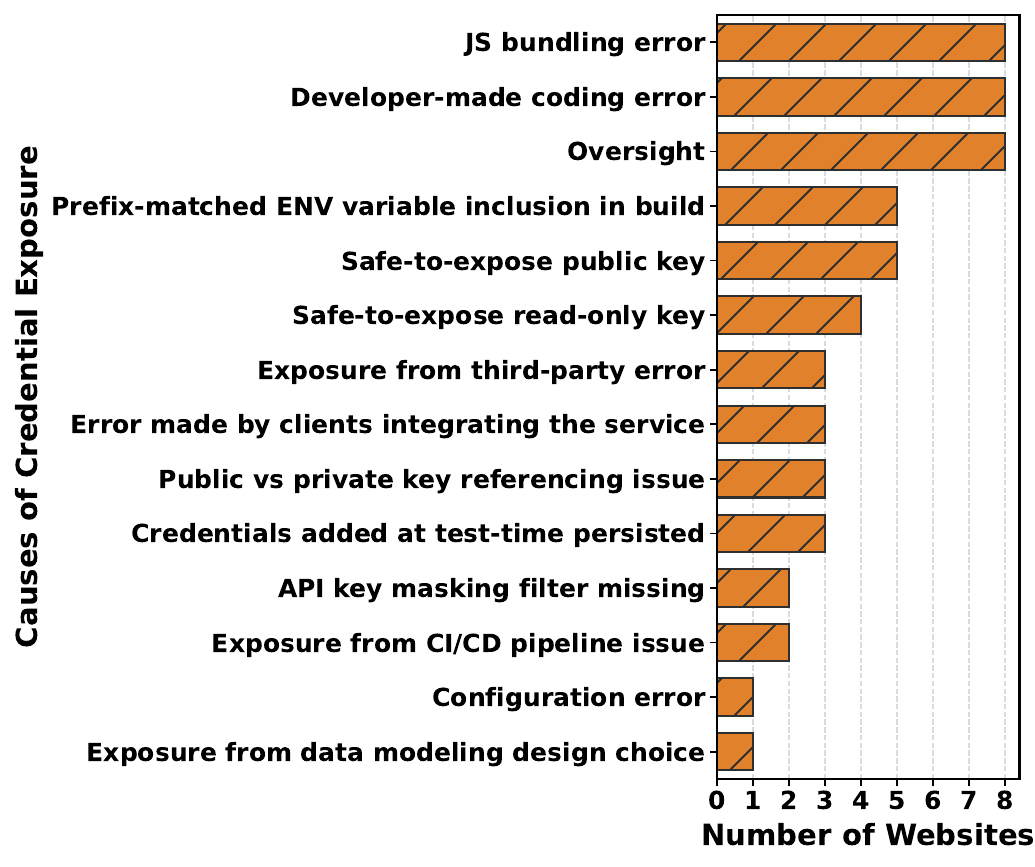}
    \vspace{-8pt}
    \caption{Exposure Causes from Disclosure Responses}
    \label{fig:cause-of-exposure}
    \vspace{-5mm}
\end{figure}

Out of 70 non-automated responses, 53~recipients acknowledged our disclosures; 39 were unaware of the exposures.
Through our disclosures, we asked for feedback to understand the cause of the reported exposure.
A total of 40 recipients provided insights about the root cause, which two authors manually analyze by assigning one or more causal categories (Figure~\ref{fig:cause-of-exposure}), with per-category inter-author agreement of Cohen's $\kappa \geq 0.92$.

\fakeparagraph{Build and Bundling Misconfigurations} \quad One of the most common causes reported by eight recipients was a JavaScript bundling issue. This correlates with our findings in \Cref{sec:exp-environment} that a significant part of the credentials are found in JavaScript bundles. %
Providers reported, as part of this bundling process, environment variables (e.g., \texttt{process.env.\textit{APIKEY}} in Anecdote~\ref{anec:nuxt}) are replaced with actual values from environment settings, resulting in referenced private credential values being embedded into public bundles that are deployed on website's server and rendered by user's browser when they visit the website.
\begin{anecquote}\label{anec:nuxt}
In our Nuxt.js (Vue) app, credentials were referenced from process.env rather than \$config, and @nuxtjs/dotenv caused all env variables---even private ones never used client-side---to be baked into the build.
\end{anecquote}
Among responses that highlighted the bundling issue (Anecdote~\ref{anec:prefix}), we coded an additional cause category, namely ``prefix-matched ENV variable inclusion in build'', that explains the inherent convention that the industry follows to prevent exposing private variables into client-side bundles.
This allows developers to use variable-naming convention prefixes such as \texttt{REACT\_APP\_} for React and \texttt{NEXT\_PUBLIC\_} or \texttt{NUXT\_PUBLIC\_} for Next.js/Nuxt.js for variable that they want to be included in browser-facing code.
However, these conventions can either be overridden by certain plugins (e.g., \textit{@nuxtjs/dotenv} described in the previous anecdote) or some configuration issue can cause all environment variable prefixes to be considered for inclusion within the bundle, resulting in exposure of even private variables and credentials.
\begin{anecquote} \label{anec:prefix}
The issue arose from relying on convention that only environment variables with a specific prefix are exposed in the build, assuming unused variables would be automatically excluded.
\end{anecquote}
Another related causal category we noticed in the responses was ``public vs. private key referencing issue'' where an incorrect key variable referencing error resulted in the exposure (Anecdote \ref{anec:public-privagte}).%
\begin{anecquote}\label{anec:public-privagte}
The exposed key was an API token intended only for our internal backend. We mistakenly referenced it as `PUBLIC\_CLOUDFLARE\_TOKEN' instead of `CLOUDFLARE\_TOKEN', causing it to be compiled into a publicly accessible script.
\end{anecquote}

\fakeparagraph{Developer Errors and Oversight} \quad  ``Developer-made coding errors'' were second most common. 
Sometimes a developer could be testing a new feature or an integration (Anecdote~\ref{anec:external}) on server side requiring private credentials and may forget to remove them once the testing concludes (``credentials added at test-time persisted'').
Alternatively, it could have been an unintentional error made by the developer that could have resulted in accidental leakage of the credentials. 
Overall, 30\% of cases (13) suggested that the exposed keys were actually ``in-use'' on the client-side, while the remaining 70\% of responses (31) described the leakage as ``accidental''. 
\begin{anecquote} \label{anec:external}
An ex-developer exposed all metadata (including credentials) through the deployment, either accidentally or for testing. The flaw carried over as the current platform was built using a copy of the existing codebase.
\end{anecquote}

Several recipients (Anecdote~\ref{anec:oversight}) described ``oversight'' as one of the reasons as well.
For instance, some businesses did not remove legacy references from their codebase even after removing the service, causing an unintended exposure from oversight stemming from several reasons---such as limited resources, limited budgets, fast-paced business culture, pressure to deliver, inexperienced or relatively new developers on the development team, coding error, and lack of or ignorance towards security best practices in lieu of speed or demonstration of value.
\begin{anecquote} \label{anec:oversight}
The exposed token was due to human error, not a bug, and the company didn't have checks in place to catch it.
\end{anecquote}

\fakeparagraph{Keys Presumed Safe to Expose} \quad Some businesses justified the exposure as being safe under two categories: ``safe-to-expose public key'' (5) and ``safe-to-expose read-only key'' (4), mainly in relation to AWS and GitHub credentials.
The presented arguments described key usage---keys with read-only scope to the blog's content, tokens with public-only scope: ``read:discussion'', and keys with restrictive IAM policy for read-only or public data ingestion (e.g., Anecdote~\ref{anec:firehouse})---to justify the exposure.
\begin{anecquote} \label{anec:firehouse}
We use the Firehose service, which doesn't allow public ingestion via unauthenticated endpoints, so we use a dedicated credential with a highly-restricted IAM policy allowing only data insertion (firehose:PutRecord) to a specific stream. We did not revoke the credentials as they are intended to be public and pose no security risk beyond a malicious actor submitting junk data or driving up costs---an inherent risk in any public ingestion system, for which we have monitoring and mitigation plans. We are aware that client-side AWS credentials aren't ideal, but a proxy server solely to sign PutRecord requests would add non-negligible costs without meaningful security benefit.
\end{anecquote}
\vspace{3pt}
However, follow-up clarification revealed that several recipients were unaware that these credentials in fact granted broader, unintended permissions. 
In one of the cases (Anecdote~\ref{anec:private-public}), client users were observed to mistakenly enter private keys in a field meant for public keys.
Our further investigation showed that even ostensibly ``safe'' public or read-only credentials can enable adversaries to flood backend services, inflate operational costs, poison downstream data, and create latent privilege-escalation risk if their scope is later extended.
Finally, we hypothesize that some organizations, concerned about reputational and compliance risks, sought to frame the reported exposure as safe by design rather than a lapse in security controls or vendor oversight, even though our disclosures were frequently followed by prompt removal or revocation of the keys. 

\begin{anecquote} \label{anec:private-public}
The exposure appears caused by users manually entering their Stripe secret keys (sk\_...) into a field meant for publishable keys (pk\_...). These keys are user-provided, not generated by our systems, indicating incorrect input rather than a system-side leak. We have since added frontend validation to prevent this, so affected entries were likely made beforehand.
\end{anecquote}

\fakeparagraph{Known and Intentional Exposures} \quad Around 14 responses indicated (e.g., Anecdote~\ref{anec:aware-not-solved}) that the recipients were aware of the exposure. There were various causes behind such intentional exposure as discussed previously such as developmental testing or misconception about all public keys to be safe to expose. 
\begin{anecquote} \label{anec:aware-not-solved}
We were aware of this issue, though it regrettably remained unaddressed.
\end{anecquote}
One of the responses revealed that the developer not only intentionally exposed the key, but also made efforts to hide the exposure to evade internet scanning tools (see Anecdote~\ref{anec:oversight-method}).
\begin{anecquote} \label{anec:oversight-method}
We implemented basic obfuscation primarily to prevent static analysis tools from flagging them. On older code we use simple Base64 encoding (in ***), while on newer code we use a string ``scramble'' (in ***). Which obfuscation did you detect?
\end{anecquote}

\section{Recommendations}
Based on our root cause analysis and disclosure feedback, we provide the following recommendations for stakeholders, annotated with the number of supporting responses ($n$).

\fakeparagraph{For Developers} \quad
Most exposed credentials originate from first-party JavaScript resources and build artifacts. We recommend:
(1) integrating credential-scanning into CI/CD for {both} source and generated bundles, failing builds when credentials are detected ($n$=20);
(2) adopting {strict} environment-variable allowlisting, where only variables explicitly prefixed for public use are included in bundles ($n$=12);
(3) enforcing static checks (e.g., pre-commit hooks) that reject common credential patterns and direct references to sensitive environment variables ($n$=10);
(4) avoiding client-side credentials by routing sensitive API calls through backend proxies ($n$=3); and
(5) using separate test credentials in non-production environments to limit the impact of accidental exposures ($n$=3).

\fakeparagraph{For Framework and Tool Maintainers} \quad
Observed credential exposure is frequently linked to bundling behavior and framework-level handling of environment variables. Frameworks and build tools can reduce this risk by:
(1) shipping secure-by-default configuration systems where developers explicitly declare which configuration entries are client-visible, treating all undeclared entries as secret and failing the build or raising a prominent warning if they are referenced in client code ($n$=12);
(2) providing build-stage exposure detection that scans bundles for high-entropy tokens and known credential formats before deployment ($n$=12); and
(3) clearly documenting and exemplifying the separation between server-only and client-visible configuration, avoiding recommended patterns that place long-lived credentials in client code ($n$=4).

\fakeparagraph{For Service Providers} \quad
Exposed credentials in our study involve cloud, payment, and messaging APIs, including cases where operators incorrectly treated ``public'' credentials as safe. Service providers can reduce this risk by:
(1) adopting default-deny permission models, in which keys start with minimal scopes and require explicit elevation for sensitive operations ($n$=22);
(2) providing integration guidance and defaults that avoid long-lived credentials in client code, instead promoting short-lived credentials, per-environment separation, and automated rotation ($n$=21); and
(3) offering clearly separated key types for client and server use, where client-side credentials are narrowly scoped, rate-limited, origin-bound, and revocable without disrupting backend workflows ($n$=8).

\section{Conclusion}
In this work, we investigated the public disclosure of sensitive API keys across the web. Despite limiting our analysis to only 14~service providers, we find that credential exposure is more common than previously believed, but only visible through dynamic analysis of production systems on the web. Our results show that the vast majority of leaks are introduced during the build process and materialize exclusively in live production environments (e.g., within JavaScript bundles), making the static scanning methods used in prior work fundamentally insufficient for the web. In many cases, exposures extend beyond online environments, enabling potential supply-chain compromise of production systems. By conducting a large-scale disclosure campaign and analyzing the responses, we identify the root causes of exposure and derive recommendations for website developers, framework and tool maintainers, and service providers. 
We hope our findings not only increase awareness but also catalyze the community to develop robust, web-native security architectures that inherently resist credential misuse.

\begin{acks}
This work was supported in part by the National
Science Foundation under Grant Number \#2540803,
as well as a Sloan Research Fellowship and a gift from Google, Inc. Georgios Smaragdakis was supported by the European Union under the Horizon Europe Programme as part of the project SafeHorizon (Grant Agreement \#101168562).
Any opinions, findings, and conclusions or recommendations expressed in this material are those of the author(s) and do
not necessarily reflect the views of the funding organizations.
\end{acks}

\bibliographystyle{ACM-Reference-Format}
\bibliography{references}

\appendix  

\section{Open Science} %
Given the sensitive nature of our findings, we release only the source code of our monitoring crawler (\url{https://github.com/stanford-esrg/crawler-keys-on-doormats}) and do not share analysis data.
\section{Ethical Considerations}
Our Institutional Review Board (IRB) determined that this research does not involve human subjects. In all phases of the study, we ensure that sensitive data (e.g., exposed credentials) are securely stored and accessed only by authorized personnel, maintaining {restricted} physical and digital access to {any} study data.
In the scanning phase, we run all measurements in a secured, dedicated environment and follow the best practices set forth by Durumeric et~al.~\cite{durumeric2024ten}. For validation, we interact only with non-intrusive service endpoints and store no raw response data, using them solely to check credential validity via status codes. We never used discovered credentials to access privileged information or resources. In the disclosure phase, we notified affected parties and sent reminders with remediation suggestions, which measurably reduced risk and were repeatedly acknowledged positively by the contacted organizations.
Similar to prior work, we could not notify all affected organizations; however, we manually identified and prioritized those in critical infrastructure sectors %
and notified them. 
While most exposed credentials were removed, we acknowledge that some residual harm is unavoidable. As with much vulnerability research, adversaries could replicate our methodology to cause damage.
Therefore, we also contacted major service providers (e.g., Azure, Amazon Web Services) to take proactive measures toward sustainably addressing this issue. 

\begin{figure}[htbp]
    \centering
    \includegraphics[width=\linewidth]{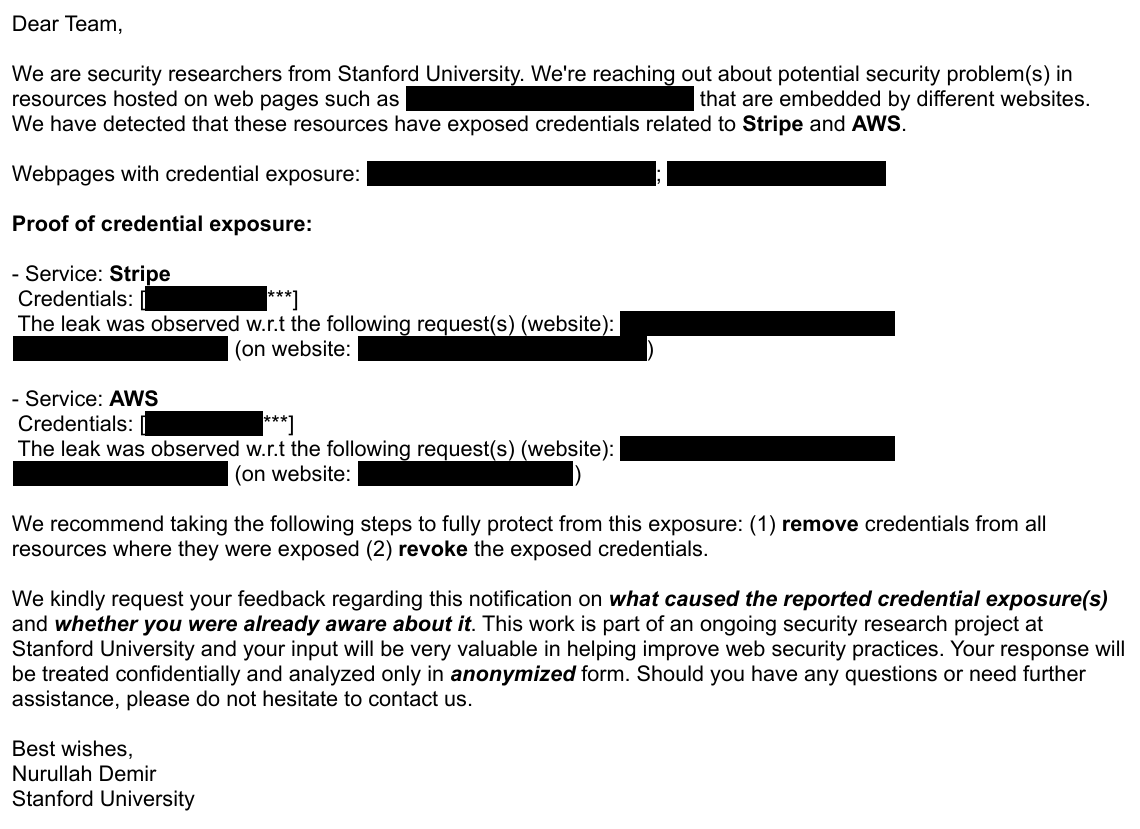}
    \caption{Email Template of Our Responsible Disclosures.}
    \label{fig:email-template}
\end{figure}

\begin{figure}[htbp]
    \centering
    \includegraphics[width=\linewidth]{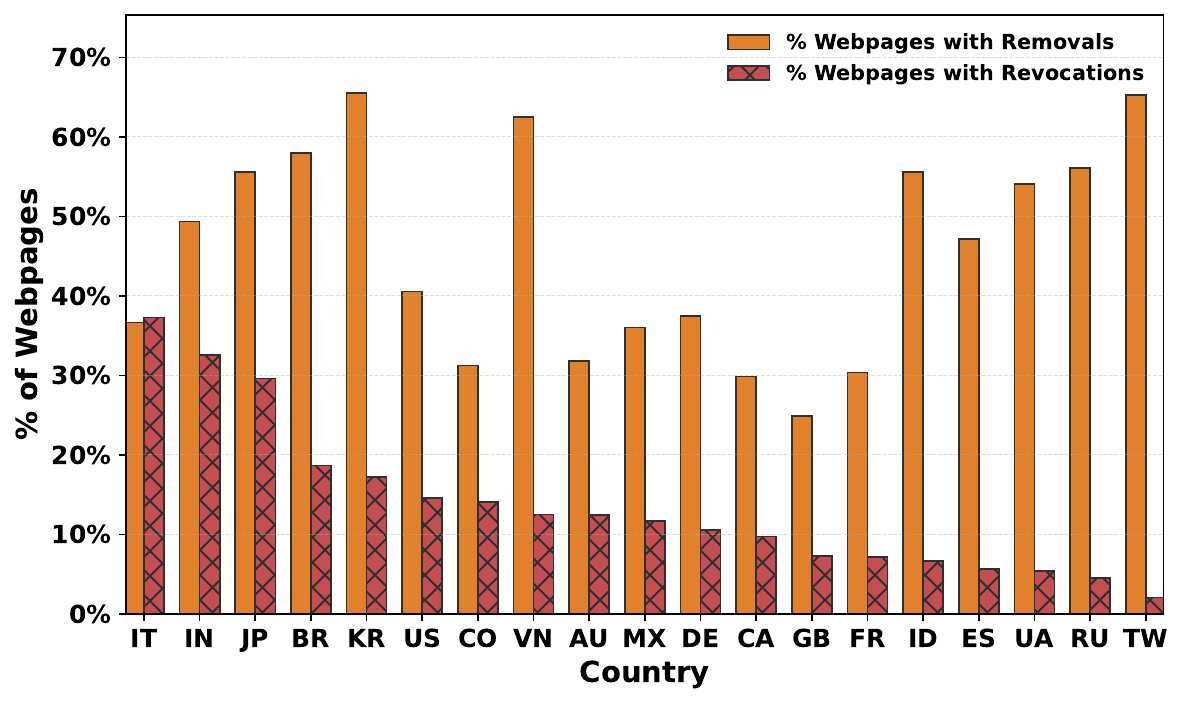}
    \caption{Aggregated Distribution of Credential Removals and Revocations per Country.}
    \label{fig:country-removals-revocations}
\end{figure}

\begin{figure}[htbp]
    \centering
    \includegraphics[width=\linewidth]{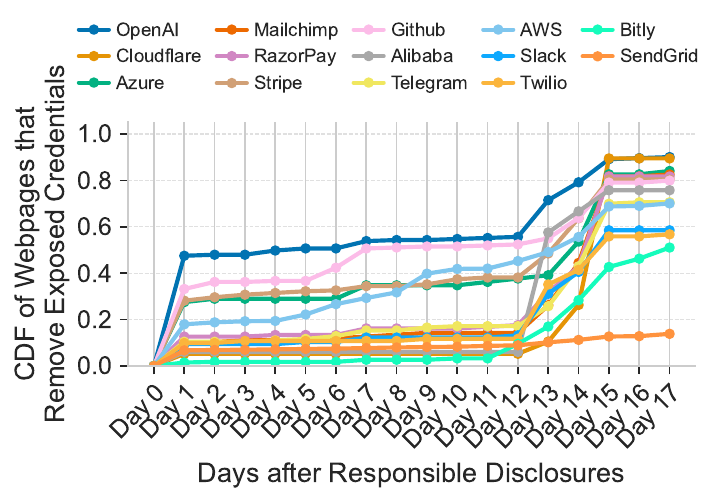}
    \caption{CDF of Websites That Remove Exposed Credentials Reported in Our Disclosures.}
    \label{fig:cdf-website-removals}
\end{figure}

\begin{figure*}[t]
  \centering
  \hspace{-3mm}
  \subfigure[By Website Rank]{
      \includegraphics[width=0.492\textwidth]{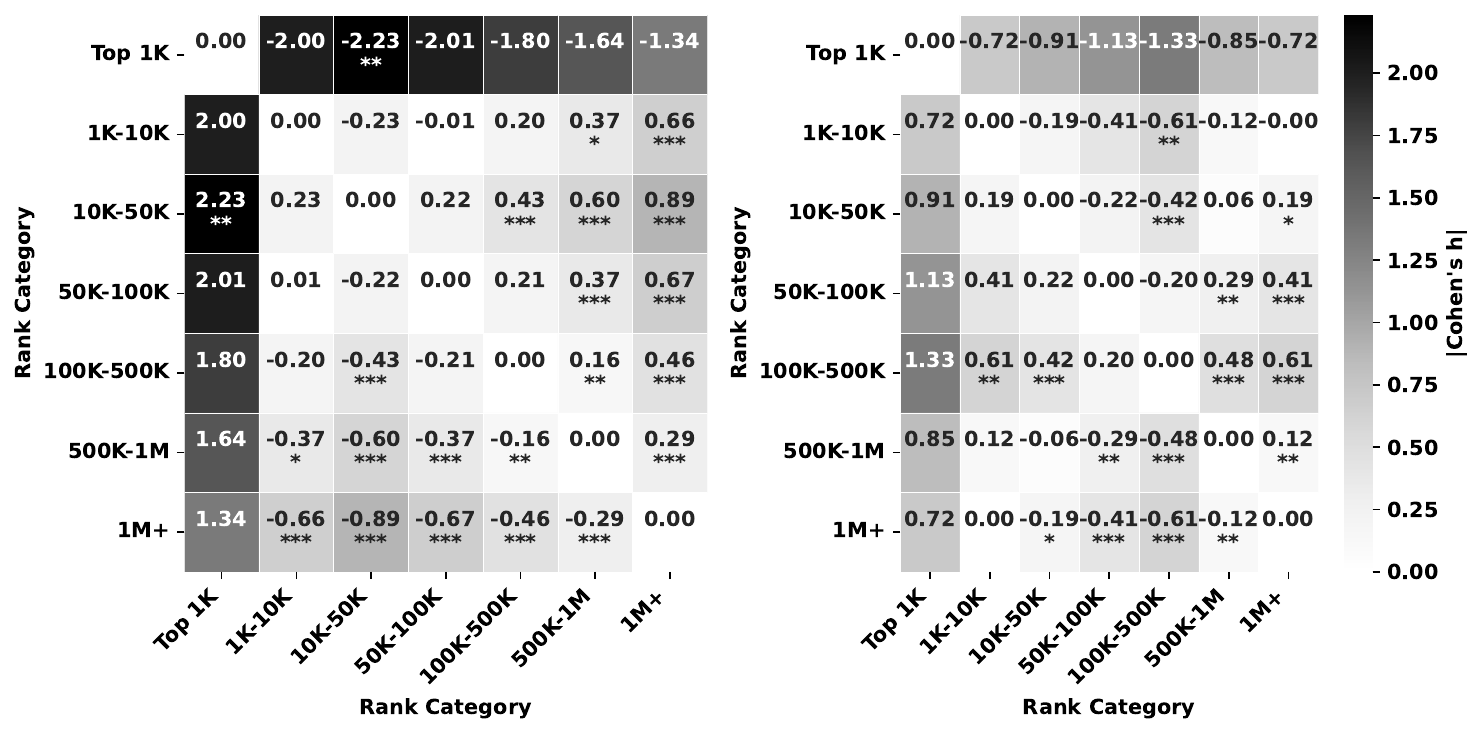}
      \label{fig:stat-ranks}
  }\hspace{-1mm}%
  \subfigure[By Website Category]{
  \includegraphics[width=0.492\textwidth]{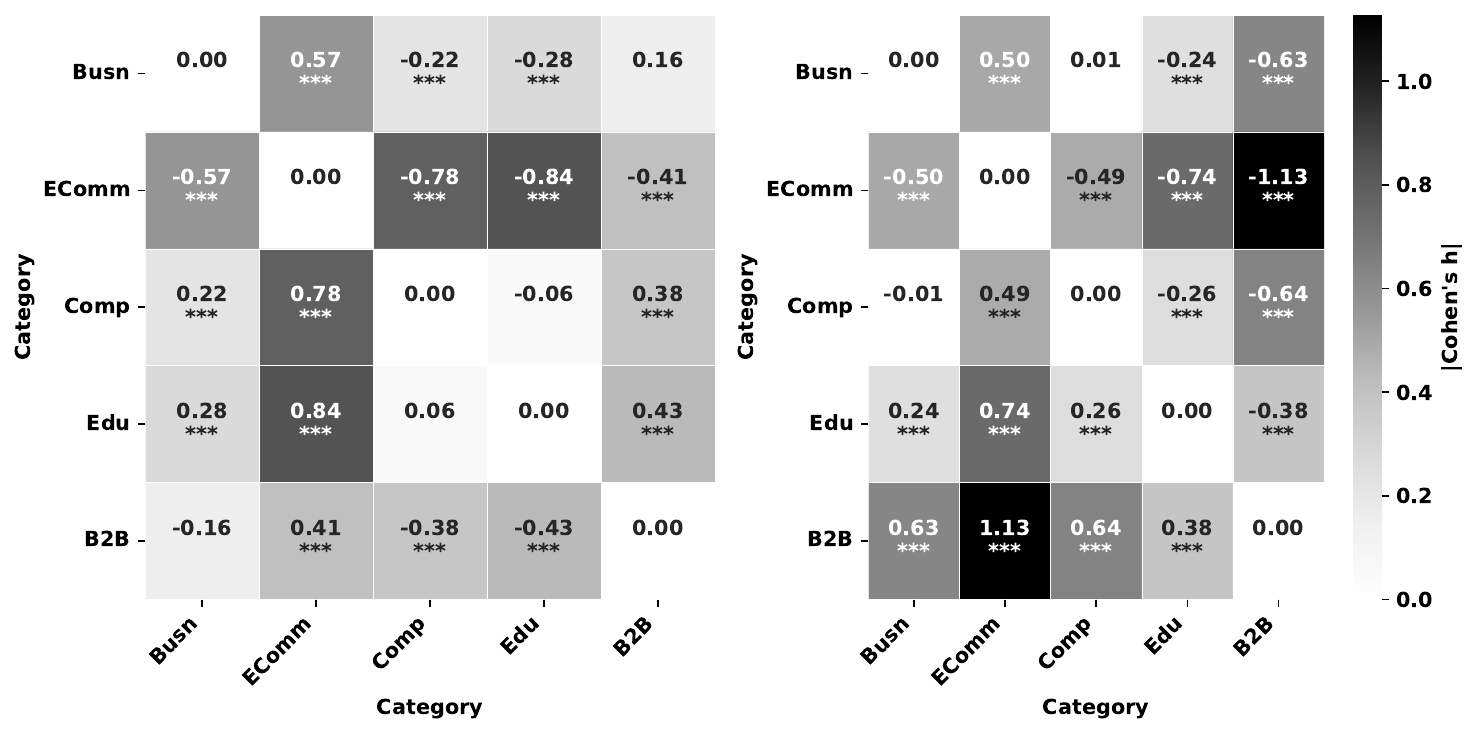}
      \label{fig:stat-cats}
  }\hspace{-3mm}
  \vspace{-5mm}
  \caption{Remediation Differences by Website Rank and Category---\textnormal{Cohen's $h$ for removal (left) and revocation (right) rates across (a) rank and (b) website categories. * $p<0.1$, ** $p<0.05$, *** $p<0.01$.}}
  \label{fig:rank-category-remediations}
  \vspace{-2mm}
\end{figure*}

\section{Disclosure and Further Findings}

Figure~\ref{fig:email-template} shows the template of a sample disclosure notification.
Figure~\ref{fig:rank-category-remediations} compares binary removal and revocation outcomes across Tranco rank categories and website categories. 
Figure~\ref{fig:cdf-website-removals} represents a CDF of remediations across different days post disclosure for all API services.
Figure~\ref{fig:country-removals-revocations} shows the fraction of websites that remove and revoke exposed credentials, broken down by country.

\section{Monitoring Crawler Details}
\label{app:craler-details}

We developed a custom browser crawler to visit webpages and collect network traffic data (i.e., requests, responses, headers, and payloads) on a daily basis to monitor credential removals.
The crawler was implemented using Puppeteer (v24.24.1) for Chrome (v142) on NodeJS (v18.19.0), following best practices of web crawling~\cite{demi2023betterweb}.
We hook onto Puppeteer's \texttt{page.on()} to persistently capture network traffic. 
Each website was visited independently in a stateless manner (with a 60-seconds load timeout) using a separate user-data directory and stealth flags (e.g. overriding \texttt{window.navigator} object)  to avoid bot detection. 
We wait 15 seconds once the page is fully loaded, to allow capturing any pending network traffic.
To crawl thousands of webpages on a daily basis, we run 20~parallel Docker containers (v28.5.1).

\clearpage

\balance

\end{document}